\documentclass[11pt,a4paper]{article}

\usepackage[a4paper,margin=2.5cm]{geometry}
\usepackage{setspace}
\usepackage[T1]{fontenc}
\usepackage[utf8]{inputenc}
\usepackage{lmodern}
\usepackage{microtype}

\usepackage{amsmath}
\usepackage{amssymb}
\usepackage{bm}
\usepackage{physics}
\usepackage{graphicx}
\usepackage{xcolor}
\usepackage{booktabs}

\newenvironment{paperabstract}
{\begin{center}
		{\bfseries\large Abstract}
	\end{center}
	\vspace{-0.6em}
	\begingroup
	\small
	\setstretch{1.08}
	\leftskip=2.0em \rightskip=2.0em
	\noindent\ignorespaces}
{\par\endgroup\vspace{1.8em}}

\usepackage[colorlinks=true,linkcolor=blue!50!black,citecolor=blue!50!black,urlcolor=blue!50!black]{hyperref}

\begin{document}
	
	\begin{center}
		\vspace*{-1.0em}
		{\LARGE\bfseries Quantum-enhanced Matterwave Interferometry\\[0.3em] in a Ring Waveguide\par}
		\vspace{2.0em}
		{\large Sriganapathy Raghav and Utpal Roy\par}
		\vspace{0.9em}
		{\small\itshape Department of Physics, Indian Institute of Technology Patna,\\
			Bihar-801103, India\par}
		\vspace{0.5em}
		{\small Corresponding author: \texttt{uroy@iitp.ac.in}\par}
	\end{center}
	
	\vspace{1.5em}
	
	\begin{paperabstract}
		We propose a quantum-enhanced matter-wave interferometry protocol based on fractional-revival dynamics of Bose--Einstein condensate in a ring, combined with a controlled interaction quench. Using the Truncated Wigner Approximation, we show that nonlinear evolution to the quarter revival generates modest spin squeezing of $-2.03\pm0.28\;$dB, which subsequently degrades under further nonlinear evolution. In contrast, starting from a non-interacting condensate, a controlled interaction quench applied after the initial quarter-revival beam-splitter operation generates substantially stronger spin squeezing of $-14.57\pm0.27\;$dB. We further show that phase control at $t=3T_r/8$ rotates the squeezed quadrature onto the phase-sensitive direction $\hat{J}_y$, producing a phase-squeezed state at $t=T_r/2$. The subsequent quarter-revival evolution converts this phase-squeezed state into a number-squeezed state, enabling population-difference readout. The resulting sequence provides a route to quantum-enhanced interferometry using the intrinsic fractional-revival dynamics of a ring together with controlled interaction and phase operations.
	\end{paperabstract}
	
\section{Introduction}

Entanglement between the atoms of an ensemble can suppress quantum fluctuations and enable phase sensitivities below the standard quantum limit (SQL), $\Delta\phi_{\mathrm{SQL}}=1/\sqrt{\mathcal{N}}$, imposed by $\mathcal{N}$ uncorrelated atoms~\cite{gross2012spin,pezze2018quantum}. Spin squeezing provides a widely used route to generate such entanglement by redistributing the quantum fluctuations of a collective pseudospin, reducing the variance along one direction below the SQL at the expense of increased fluctuations along the conjugate direction~\cite{kitagawa1993squeezed,Wineland1992}. When the squeezed quadrature is aligned with the interferometric readout, this leads directly to enhanced phase sensitivity.

Interaction-induced one-axis twisting (OAT) has enabled metrologically useful spin squeezing in Bose-Einstein condensates (BECs), using either internal atomic states in a binary condensate~\cite{Gross2010,Riedel2010} or spatially separated modes of a single-component condensate~\cite{Esteve2008}. In the single-component case, early experiments reached $-2.3$ dB and $-3.8$ dB of squeezing by splitting a condensate across lattice sites~\cite{Esteve2008}. This was later improved to $-7.8(1)$ dB in an integrated BEC interferometer~\cite{Berrada2013}, and most recently to close to $-10$ dB using the oscillatory squeezing dynamics of a multimode bosonic Josephson junction~\cite{Zhang2024}. Theory predicts that even larger squeezing is possible with optimally engineered spatial-mode dynamics: up to $-18$ dB through parametric-squeezing amplification~\cite{Jager2015}. Together, these results show that single-component condensates are a viable and steadily improving platform for generating nonclassical states, and that the achievable squeezing depends strongly on the mechanism used to prepare it.

These mechanisms come with different experimental trade-offs. Light-mediated squeezing, using cavity feedback or quantum nondemolition measurement, can achieve substantial spin squeezing~\cite{Leroux2010}. However, the resulting squeezed states can be sensitive to atomic motion and other mechanisms that limit their useful lifetime~\cite{Leroux2010b}, while cavity-based implementations additionally require stable atom--cavity coupling and optical probing~\cite{Hosten2016}. Collisional OAT avoids this optical overhead. However, the same interactions that generate the squeezing can also induce spatial dynamics, leading to multimode evolution that may degrade the mode matching required for high-contrast interferometric recombination~\cite{Haine2014}. Thus, existing approaches involve different compromises between optical complexity, spatial-mode dynamics, and interferometric control. This is why alternative schemes remain valuable: ones where both the interferometric operations and the state preparation come predominantly from the intrinsic evolution of the matter wave itself. 

Ring confinement offers exactly this kind of setting, through the revival and fractional-revival (FR) dynamics of a localized matter wave. The quantized angular-momentum modes of a ring disperse quadratically, so an initially localized wavepacket rephases periodically. At fractional revival times, it reconstructs into a coherent superposition of spatially separated daughter wave packets~\cite{bera2020matterwave,kialka2020orbital,raghav2024dispersion}. At the quarter revival in particular, the wavepacket splits into two daughter clouds, which recombine at the half revival. This gives the splitting and recombination needed for matter-wave interferometry, using nothing but the free evolution~\cite{kialka2020orbital,raghav2024dispersion}. Interactions modify this revival structure by adding a nonlinear energy scale~\cite{raghav2025nonlinearity}. As in other nonlinear matter-wave interferometers, they also bring quantum fluctuations that can introduce phase noise and depletion, degrading the coherence of the split condensate~\cite{martin2012quantum,haine2018quantum}.

In this work, we develop a quantum-enhanced interferometric protocol based on FR dynamics in a single-component BEC confined to a ring, without optical beam splitters. We show that phase control at the one-eighth revival allows the quarter revival to function as a generalized beam splitter. We then investigate the effect of interactions and quantum fluctuations on the FR dynamics using the Truncated-Wigner Approximation (TWA) and characterize the coherence of the split condensates. Finally, we introduce a controlled interaction quench after the initial beam-splitter operation to generate spin squeezing, followed by FR recombination. By optimizing the interaction strength and quench duration, we obtain spin squeezing of $-14.57\pm0.27$~dB and demonstrate quantum-enhanced interferometric performance using the FR dynamics of the ring-confined condensate.

The paper is organized as follows. Section~2 introduces the fractional-revival dynamics of a non-interacting BEC in a ring, with emphasis on the quarter revival as a generalized beam splitter, and then examines the effect of repulsive interactions. Section~3 incorporates quantum fluctuations using the TWA and discusses condensate fraction, phase coherence, and the implications for FR interferometry. Section~4 presents the quantum-enhanced interferometry protocol based on a controlled interaction quench and discusses the resulting squeezing and its robustness. Section~5 concludes the paper.

\begin{figure*}[th]
	\centering
	\includegraphics[width=16 cm]{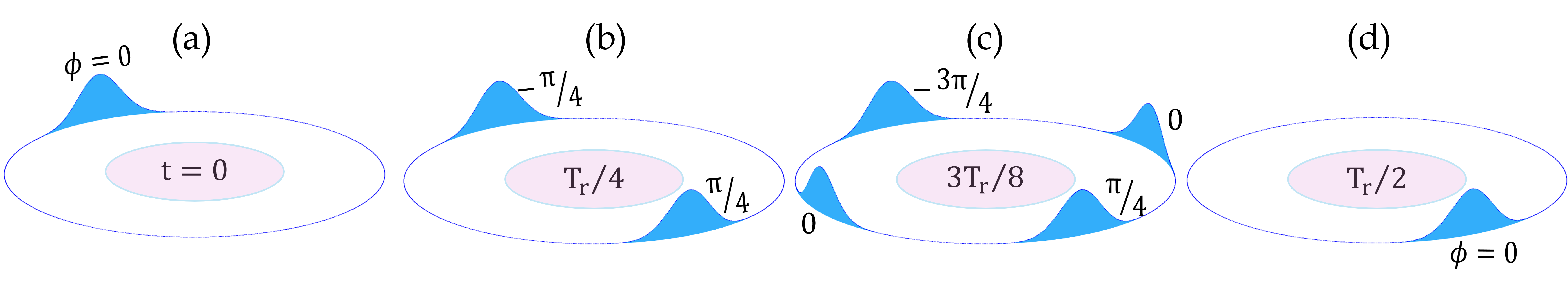}
	\caption{The probability density and the corresponding phase, for a non-interacting BEC, for fractional revival instances (a) $t=0$, (b) $t=\frac{T_{\mathrm{r}}}{4}$, (c) $t=\frac{3T_{\mathrm{r}}}{8}$, and (d) $t=\frac{T_{\mathrm{r}}}{2}$. Here, $T_r=\frac{4\pi m R^2}{\hbar}$ is the revival time as given in Eq.\eqref{qrev}.}
	\label{FR_phase_den_NI}
\end{figure*}
 
\section{Fractional Revivals in a Ring}\label{s3_c7}
A localized wavepacket, when released in a ring, exhibits fractional revival (FR) dynamics~\cite{averbukh1989fractional,robinett2004quantum}. The initial wavepacket disperses during propagation and partially recombines at specific instances, forming a coherent superposition of spatially separated copies of the initial cloud. At the full revival time, the initial cloud is reconstructed with the same spatial profile and position. In a non-interacting BEC, the occurrence of the revival of the initial cloud is purely due to the quadratic dependence of single-particle energies on the angular quantum number $l$ in a ring of radius $R$, given by $E_{l}=\frac{\hbar^2l^2}{2mR^2}$. This quadratic dependence results in periodic rephasing of the Fourier components, leading to a full revival at~\cite{bera2020matterwave,kialka2020orbital}
\begin{equation}\label{qrev}
	T_{\mathrm r}=\frac{4\pi m R^2}{\hbar},
\end{equation}
At rational fractions of the revival time, the FR instances occur at
\begin{equation}
	t=\frac{p}{q}T_{\mathrm r},
\end{equation}
where $p$ and $q$ are coprime integers, i.e., $\text{gcd}(p,q)=1$.

The fractional-revival copies of the initial condensate are a coherent superposition of distinct angular momentum eigenstates. In a 1D ring of circumference $L=2\pi R$, the initial non-interacting matter-wavepacket can be expressed in terms of the eigenmodes of the ring as
\begin{equation}
	\psi(x,0)=\sum_{l\in\mathbb{Z}}c_{l}\;e^{ilx/R},\quad c_{l}=\frac{1}{\sqrt{2\pi R}}\int\limits_{-\pi R}^{\pi R}\text{d}x\;\psi(x,0)\;e^{-ilx/R}.
\end{equation}
The state at a FR instance becomes
\begin{equation}\label{psi_@t}
	\psi\big(x,pT_{\mathrm r}/q\big)=\sum_{l\in\mathbb{Z}}c_{l}e^{-i2\pi pl^2/q}e^{ilx/R}.
\end{equation}
Here, for each $l\in\mathbb{Z}$, there is a unique decomposition
$l=aq+b$, with $a\in\mathbb{Z}$ and $b\in\{0,\dots,q-1\}$.
Consequently, the quadratic phase factor is periodic in $l$ modulo $q$
and can be expressed as a finite Fourier series ~\cite{bigourd2008factorization,wolk2011factorization,lakshmibala2022revivals}:
\begin{equation}\label{Gauss_sum}
	e^{-i2\pi pl^2/q}=\frac{1}{q}\sum_{k=0}^{q-1}A_{k}(p,q)\;e^{-i2\pi lk/q},
\end{equation}
where $A_{k}$ are the quadratic Gauss coefficients given by:
\begin{equation}
	A_{k}(p,q)=\sum_{b=0}^{q-1}e^{-i2\pi pb^2/q}\;e^{i2\pi bk/q}.
\end{equation}
Substituting Eq.~\eqref{Gauss_sum} in Eq.~\eqref{psi_@t} and rearranging, we get
\begin{equation}\label{CS_FR}
	\psi\big(x,pT_{\mathrm r}/q\big)=\frac{1}{q}\sum_{k=0}^{q-1}A_{k}(p,q)\;\psi\big(x-2\pi Rk/q,0\big).
\end{equation}

Equation~\eqref{CS_FR} shows that, at each FR instance $t=pT_{\mathrm r}/q$, the wavepacket is a coherent superposition of translated copies of the initial BEC, with their relative amplitudes and phases determined by the Gauss coefficients $A_{k}/q$. Figure~\ref{FR_phase_den_NI}(a)-(d) shows the density and phase of the wavepacket at the FR instances $t=0$, $t=T_{\mathrm r}/4$, $t=3T_{\mathrm r}/8$, and $t=T_{\mathrm r}/2$, respectively. At $t=T_{\mathrm r}/4$, the initial cloud is split into two spatially separated copies located at opposite points of the ring, with a relative phase of $\pi/2$. At $t=3T_{\mathrm r}/8$, four spatially separated copies are formed, while at $t=T_{\mathrm r}/2$, the initial cloud is reconstructed at the diametrically opposite point of the ring with the same phase. These particular FR instances are used below to construct a generalized beam splitter with a continuously tunable splitting ratio.

\subsection{Quarter-revival as a Generalised Beam-splitter:}
\label{UnitOp}

The quarter revival at $t=T_{\mathrm r}/4$ naturally acts as a balanced beam splitter, producing two spatially separated clouds with equal populations. To obtain a continuously tunable splitting ratio, we exploit the four-cloud structure at $T_{\mathrm r}/8$. The key idea is to use two successive eighth revivals, with a phase imprint applied to the two auxiliary fractional-revival clouds between them \cite{li2009local}. At $t=T_{\mathrm r}/8$, the phase acquired by the $l$th angular momentum component is $e^{-i2\pi l^2/8}=e^{-i\pi l^2/4}$. This phase has a period of four in $l$, since $e^{-i\pi(l+4)^2/4}=e^{-i\pi l^2/4}$.

Therefore, the wavepacket at $t=T_{\mathrm r}/8$ can be expressed as a coherent superposition of four copies of the initial wavepacket located at angular positions separated by $\pi/2$. Writing $\epsilon=e^{-i\pi/4}$, the corresponding amplitudes are
\begin{equation}
	c_0=\frac{\epsilon}{2},
	\qquad
	c_1=\frac{1}{2},
	\qquad
	c_2=-\frac{\epsilon}{2},
	\qquad
	c_3=\frac{1}{2}.
\end{equation}
Thus, each of the four fractional-revival clouds carries one quarter of the total population, while their relative phases are fixed by the quadratic phase evolution. We define the four-cloud basis as
\begin{equation}
	\mathcal{B}_4=
	\left\{
	|0\rangle,
	|\pi/2\rangle,
	|\pi\rangle,
	|-\pi/2\rangle
	\right\}.
\end{equation}
Since the ring Hamiltonian commutes with spatial translations, a wavepacket initially localized at any of these four positions evolves according to the same transformation, translated around the ring. The corresponding $T_{\mathrm r}/8$ propagator in this basis is
\begin{equation}
	U_1
	=
	\frac{1}{2}
	\begin{pmatrix}
		\epsilon & 1 & -\epsilon & 1\\
		1 & \epsilon & 1 & -\epsilon\\
		-\epsilon & 1 & \epsilon & 1\\
		1 & -\epsilon & 1 & \epsilon
	\end{pmatrix},
\end{equation}
In particular,
\begin{equation}
	U_1|0\rangle
	=
	\frac{1}{2}
	\begin{pmatrix}
		\epsilon\\
		1\\
		-\epsilon\\
		1
	\end{pmatrix},
\end{equation}
showing that an initially localized cloud is split into four equally populated fractional-revival clouds.

We now identify the two opposite clouds $|0\rangle$ and $|\pi\rangle$ as the two logical modes of the beam splitter. The clouds $|\pi/2\rangle$ and $|-\pi/2\rangle$ serve as auxiliary modes during the intermediate $T_{\mathrm r}/8$ evolution. A phase $\alpha$ applied to these two auxiliary clouds is represented by
\begin{equation}
	P(\alpha)
	=
	\operatorname{diag}
	\left(
	1,e^{i\alpha},1,e^{i\alpha}
	\right).
\end{equation}
The generalized beam splitter is then given by the composite operation
\begin{equation}
	U_{\mathrm{BS}}(\alpha)
	=
	U_1P(\alpha)U_1. \label{GBS}
\end{equation}
Applying this operation to the two logical modes gives
\begin{align}
	U_{\mathrm{BS}}(\alpha)|0\rangle
	&=
	\frac{e^{i\alpha}-i}{2}|0\rangle
	+
	\frac{e^{i\alpha}+i}{2}|\pi\rangle,
	\\
	U_{\mathrm{BS}}(\alpha)|\pi\rangle
	&=
	\frac{e^{i\alpha}+i}{2}|0\rangle
	+
	\frac{e^{i\alpha}-i}{2}|\pi\rangle.
\end{align}
The amplitudes of the auxiliary clouds $|\pi/2\rangle$ and $|-\pi/2\rangle$ vanish after the second one-eighth revival. Hence, the composite operation maps the two-mode subspace spanned by $|0\rangle$ and $|\pi\rangle$ back onto itself. In the ordered basis $(|0\rangle,|\pi\rangle)$, the corresponding two-mode propagator is
\begin{equation}
	U_{\mathrm{BS}}^{(2)}(\alpha)
	=
	\frac{1}{2}
	\begin{pmatrix}
		e^{i\alpha}-i & e^{i\alpha}+i\\
		e^{i\alpha}+i & e^{i\alpha}-i
	\end{pmatrix}.
\end{equation}

The matrix elements can be written as
\begin{align}
	\frac{e^{i\alpha}-i}{2}
	&=
	e^{i(\alpha/2-\pi/4)}
	\cos\left(\frac{\alpha}{2}+\frac{\pi}{4}\right),
	\\
	\frac{e^{i\alpha}+i}{2}
	&=
	i e^{i(\alpha/2-\pi/4)}
	\sin\left(\frac{\alpha}{2}+\frac{\pi}{4}\right).
\end{align}
Therefore, defining
\begin{equation}
	R_x(\theta)
	=
	e^{-i\theta\sigma_x^{(\mathrm L)}/2},
\end{equation}
where
\begin{equation}
	\sigma_x^{(\mathrm L)}
	=
	|0\rangle\langle\pi|
	+
	|\pi\rangle\langle0|,
\end{equation}
we obtain
\begin{equation}
	U_{\mathrm{BS}}^{(2)}(\alpha)
	=
	e^{i(\alpha/2-\pi/4)}
	R_x\left(-\frac{\pi}{2}-\alpha\right).
\end{equation}
Thus, apart from an overall phase, the composite fractional-revival operation corresponds to an $x$-rotation with a continuously tunable angle $\theta(\alpha)=-\frac{\pi}{2}-\alpha$.

The population splitting produced by the generalized beam splitter can be obtained directly from the above transformation. For an atom initially in $|0\rangle$, the populations in the two output modes are
\begin{equation}
	P_0
	=
	\left|
	\frac{e^{i\alpha}-i}{2}
	\right|^2
	=
	\cos^2\left(\frac{\pi}{4}+\frac{\alpha}{2}\right),
\end{equation}
and
\begin{equation}
	P_\pi
	=
	\left|
	\frac{e^{i\alpha}+i}{2}
	\right|^2
	=
	\sin^2\left(\frac{\pi}{4}+\frac{\alpha}{2}\right).
\end{equation}
Hence, the splitting ratio can be continuously controlled by varying the phase imprint $\alpha$. For $\alpha=0$, the phase imprint is absent and $P_0=P_\pi=\frac{1}{2}$, corresponding to the conventional balanced beam splitter. In this case, $	U_{\mathrm{BS}}(0)=U_1^2=U_2$,
where $U_2$ is the quarter-revival propagator. At $\alpha=\pm\pi/2$, the operation gives complete transfer into one of the two output modes. Thus, the conventional balanced quarter-revival beam splitter is recovered as the $\alpha=0$ member of a continuously tunable family of beam splitters.

\begin{figure*}[ht!]
	\centering
	\includegraphics[width=16 cm]{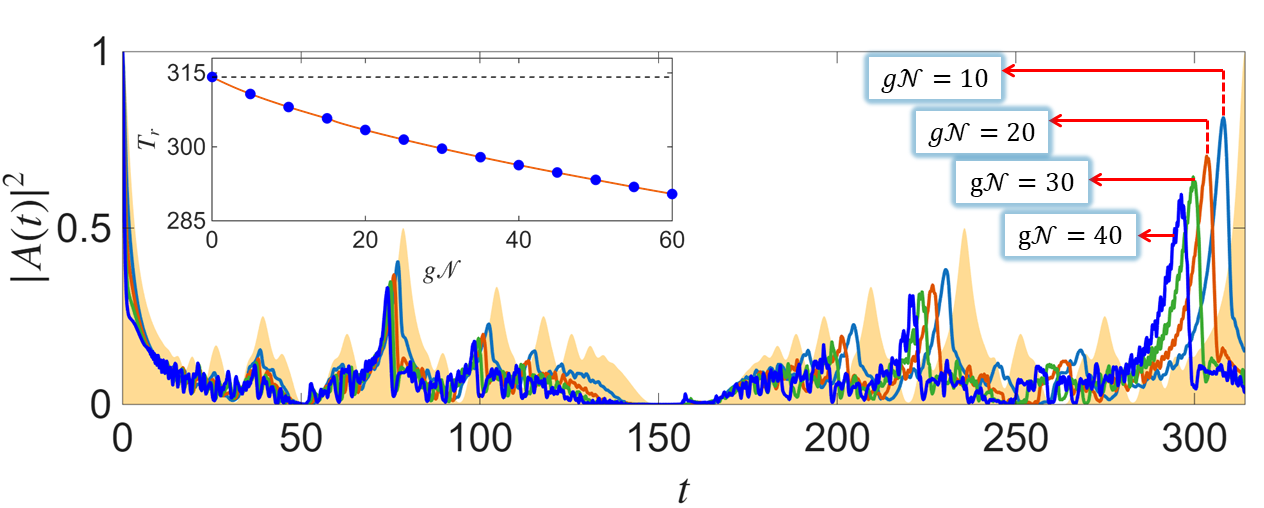}
	\caption{Survival function $|A(t)|^2$ for different nonlinearities, $\mathrm{g}\mathcal{N}=0,\;10,\;20,\;30,\;$and $40$. The non-interacting case, $\mathrm{g}\mathcal{N}=0$, is shown as a filled plot in the background. The inset plot shows the dependence of the revival time $T_{r}$ on nonlinearity, where the red-dashed line shows the revival time corresponding to a non-interacting BEC. Time is in the units of $\omega^{-1}_{x}=3.18\;$ms.}
	\label{AC}
\end{figure*}

\subsection{In the presence of repulsive interactions:}
In the presence of repulsive interactions, the nonlinear mean-field term couples different angular momentum components, modifying their relative phase evolution during propagation. Consequently, the exact quadratic phase relationship of Eq.~\eqref{Gauss_sum}, responsible for the periodic rephasing in Eq.~\eqref{CS_FR}, is altered as the interaction strength is increased. The fractional-revival structure is therefore modified from the ideal non-interacting case \cite{herrera2012echoes,raghav2025nonlinearity}.

To study the effect of repulsive interactions on the FR dynamics of the BEC in a ring, we employ the usual mean-field approach. Here, the dynamics of the matter wave is governed by the three-dimensional Gross--Pitaevskii equation~\cite{pethick2002bose,dalfovo1999theory}:
\begin{equation}
	i\hbar\partial_t \Psi(\mathbf{r},t)=\bigg[-\frac{\hbar^2}{2m}\nabla^2+V(\mathbf{r})+g_{3\text{D}}|\Psi(\mathbf{r},t)|^2\bigg]\Psi(\mathbf{r},t)
\end{equation}
where, $g_{3\text{D}}=\frac{4\pi\hbar^2a_{s}}{m}$ is the interaction strength with s-wave scattering length $a_{s}$. The external potential is $V(\mathbf{r})=\frac{1}{2}m(\omega^2_{x}x^2+\omega^2_{y}y^2+\omega^2_{z}z^2)$, such that $\omega_{y}=\omega_{z}\gg\omega_{x}$. We introduce the aspect ratio $\lambda=\frac{\omega_x}{\omega_r}$, where $\omega_{r}=\omega_{y}=\omega_{z}$ and $\omega_{x}$ are the radial and axial trap frequencies, respectively. 

For a quasi-1D cigar-shaped BEC, $\lambda\ll 1$. Consequently, in the radial direction, the BEC wavefunction is approximated by a 2D Gaussian function of width $a_{r}=\sqrt{\frac{\hbar}{m\omega_r}}$. Substituting this into the 3D GPE and integrating out the radial component, we obtain the effective quasi-1D GPE. The quasi-1D GPE is written after nondimensionalization of the axial coordinate, time and energy as $\bar{x}=x/a_{x}$, $\bar{t}=\omega_{x}t$ and $\bar{E}=E/\hbar\omega_{x}$, respectively, where, $a_{x}=\sqrt{\frac{\hbar}{m\omega_{x}}}$ is the axial harmonic oscillator length. Throughout this work, we use dimensionless variables and omit the overbars for simplicity. The resulting dimensionless quasi-1D GPE becomes~\cite{salasnich2002effective,salasnich2007matter,muruganandam2009fortran}:
\begin{equation}\label{Q1_GPE}
	i\partial_t \psi(x,t)=\bigg[-\frac{1}{2}\partial_{xx}+\frac{1}{2}x^2+\mathrm{g}|\psi(x,t)|^2\bigg]\psi(x,t) 
\end{equation}
Here, the 1D nonlinearity $\mathrm{g}=\frac{2a_{s}}{\lambda a_{x}}$, and the 1D wavefunction $\psi(x,t)$ is normalized to the total number of atoms, i.e., $\int\text{d}x|\psi(x,t)|^2=\mathcal{N}$. The harmonically trapped BEC is transferred into a quasi-1D ring, which is modelled along the $x$-direction with periodic boundary conditions $x\in[-\pi R,\pi R)$, where $R=5$ (units of $a_x$) is the ring radius. The ground state obtained by solving Eq.~\eqref{Q1_GPE} is then evolved in a 1D ring using real-time propagation to obtain the fractional revivals.

The interaction-induced change in the revival dynamics can be characterized using the survival function, which is the absolute-squared autocorrelation function~\cite{robinett2004quantum,bera2020matterwave},
\begin{equation}
	|A(t)|^2=\left|\int_{-\pi R}^{\pi R}\psi^{*}(x,0)\psi(x,t)\,\mathrm{d}x\right|^{2},
\end{equation}
shown in Fig.~\ref{AC}. The characteristic fractional-revival and full-revival peaks are observed to shift towards shorter times with increasing nonlinearity. The inset of Fig.~\ref{AC} shows the corresponding variation in the revival time $T_{\mathrm r}$ with increasing nonlinearity $\mathrm{g}\mathcal{N}$. The dashed line corresponds to the revival time in the absence of interactions for the ring radius $R=5$.

Thus, while interactions modify the precise timing and structure of the fractional revivals, the revival features remain clearly identifiable over the interaction strengths considered here. This interaction-dependent modification of the FR dynamics is taken into account in the subsequent analysis. Further, to study the effects of quantum fluctuations along with atomic interactions, we employ the Truncated Wigner Approximation in the next section.

\section{Effect of Quantum Fluctuations:}\label{s4_c7}
The Truncated Wigner Approximation (TWA) provides a beyond-mean-field description of interacting Bose gases by incorporating quantum fluctuations through a stochastic phase-space representation~\cite{steel1998dynamical,sinatra2002truncated,blakie2008dynamics}. In this approach, the quantum field is represented by an ensemble of classical fields whose initial fluctuations are sampled from the corresponding Wigner distribution and subsequently evolved according to the Gross--Pitaevskii equation. For interacting trapped condensates, the initial Wigner distribution can be constructed using Bogoliubov quasiparticle modes~\cite{sinatra2001classical,sinatra2002truncated}. Here, we use this approach to investigate the effects of quantum fluctuations on phase coherence and quantum depletion during the FR dynamics. For each realization $i$,the initial classical field is sampled as:
\begin{equation}\label{WignerSam}
	\psi_{\mathcal{C},i}(x,0)=\psi_{0}(x)+\sum_{j=1}^{M}\left(\beta_{ij}u_{j}(x)+\beta^{*}_{ij}v_{j}^{*}(x)\right)
\end{equation}
Here, $\psi_{0}(x)$ represents the ground state corresponding to Eq.~\eqref{Q1_GPE} and $\{u_{j}(x),v_{j}(x)\}$ are the Bogoliubov modes obtained by solving the Bogoliubov-de-Gennes eigenvalue problem~\cite{sinatra2001classical,pethick2002bose}. The condensate ground state is obtained by imaginary-time propagation and is subsequently used to construct the BdG matrix,
\begin{equation}
	\begin{pmatrix}
		\mathcal{L}(x) & g\psi_{0}(x)^2 \\ -g\psi^{*}_{0}(x)^2 & -\mathcal{L}^{*}(x)
	\end{pmatrix} 
	\begin{pmatrix}
		u_{j}(x)\\ v_{j}(x)
	\end{pmatrix} 
	=\epsilon_{j} 	
	\begin{pmatrix}
		u_{j}(x)\\ v_{j}(x)
	\end{pmatrix},
\end{equation} 
where $\mathcal{L}(x)=-\frac{1}{2}\partial_{xx}+\frac{1}{2}x^2+2g|\psi_{0}(x)|^2-\mu$ and $\epsilon_{j}$ represents the Bogoliubov energies. We solve the eigenvalue problem for $M$ modes using IRAM implemented in ARPACK \cite{lehoucq1998arpack}. The modes are normalized according to the Bogoliubov metric,
\begin{equation}
	\int (|u_{j}(x)|^2-|v_{j}(x)|^2)\;\text{d}x=1
\end{equation}
with the corresponding symplectic orthogonality relations between distinct quasiparticle modes~\cite{pethick2002bose}.
The stochastic amplitudes $\beta_{ij}$ are independent complex Gaussian variables satisfying, $\langle\beta_{ij}\rangle=0$ and $\langle\beta^{*}_{ij}\beta_{ik}\rangle=\frac{1}{2}\delta_{jk}$, such that each Bogoliubov mode contains, on average, one-half quantum of Wigner vacuum noise. This procedure samples the zero-temperature quantum fluctuations of the condensate within the Bogoliubov approximation~\cite{sinatra2001classical,sinatra2002truncated,blakie2008dynamics}. 
\begin{figure*}[ht!]
	\centering
	\includegraphics[width=16 cm]{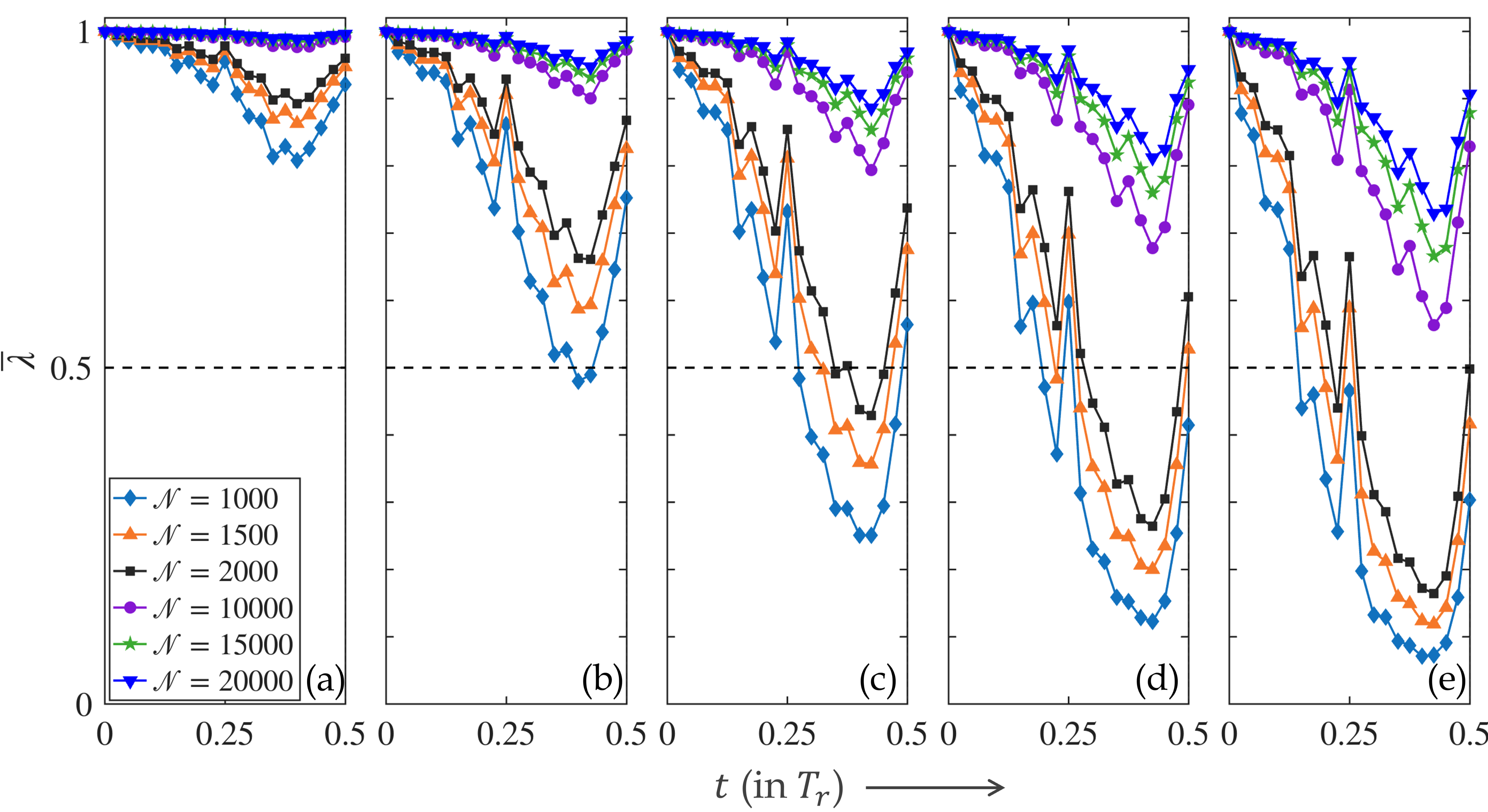}
	\caption{Condensate fraction as a function of time for different nonlinearities: (a) $\mathrm{g}\mathcal{N}=10$, (b) $\mathrm{g}\mathcal{N}=20$, (c) $\mathrm{g}\mathcal{N}=30$, (d) $\mathrm{g}\mathcal{N}=40$, and (e) $\mathrm{g}\mathcal{N}=50$. For each nonlinearity, the condensate fraction is shown for atom numbers $\mathcal{N}=1000,\;1500,\;2000,\;10000,\;15000,\;20000$. The black-dashed line indicates a condensate fraction of $50\%$. Here, $T_r$ denotes the revival time corresponding to the respective nonlinearity $\mathrm{g}\mathcal{N}$ and therefore has a different value in each panel. $T_r$ is given in units of $\omega_x^{-1}=3.18\;$ms.}
	\label{cond_fr}
\end{figure*}
\begin{figure*}[ht!]
	\centering
	\includegraphics[width=16 cm]{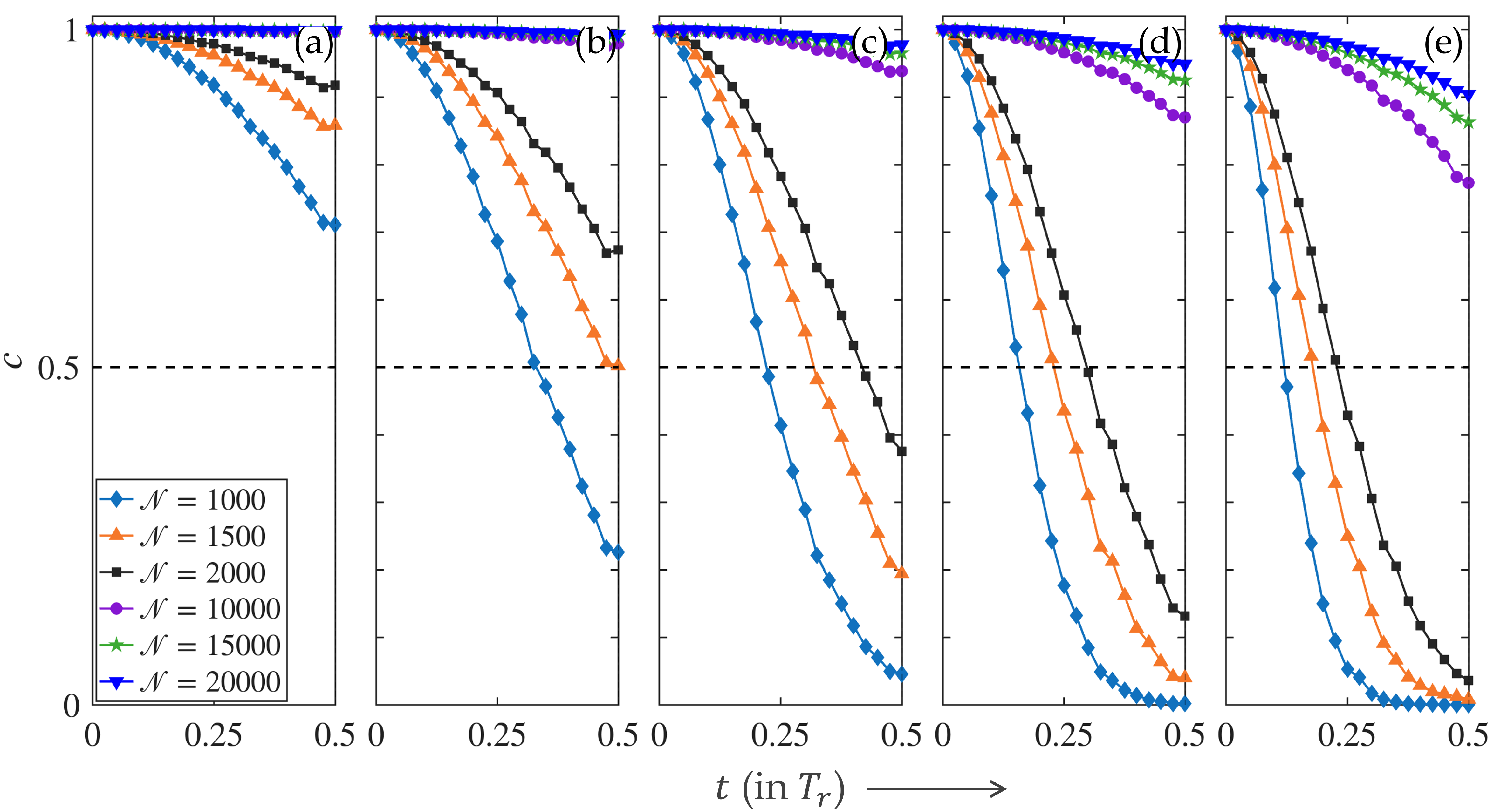}
	\caption{Phase coherence as a function of time for different nonlinearities: (a) $\mathrm{g}\mathcal{N}=10$, (b) $\mathrm{g}\mathcal{N}=20$, (c) $\mathrm{g}\mathcal{N}=30$, (d) $\mathrm{g}\mathcal{N}=40$, and (e) $\mathrm{g}\mathcal{N}=50$. For each nonlinearity, $c(t)$ is shown for atom numbers $\mathcal{N}=1000,\;1500,\;2000,\;10000,\;15000,\;20000$. The black-dashed line indicates a phase coherence of $50\%$. Here, $T_r$ denotes the revival time corresponding to the respective nonlinearity $\mathrm{g}\mathcal{N}$ and therefore has a different value in each panel. $T_r$ is given in units of $\omega_x^{-1}=3.18\;$ms.}
	\label{phase_coh}
\end{figure*}
\subsection{Ensemble dephasing and Quantum Depletion:}

The deterministic GPE describes the evolution of a single condensate wavefunction and therefore cannot capture the loss of coherence between different stochastic realizations or the transfer of atoms from the condensate into excited modes due to quantum fluctuations. To quantify these effects, we characterize the TWA ensemble using the single-particle density matrix and the phase coherence of individual stochastic trajectories.

The condensate fraction is determined using the Penrose--Onsager criterion~\cite{penrose1956bose}. The single-particle density matrix is defined in terms of the field operator as
\begin{equation}
	\rho^{(1)}(x,x';t)
	=
	\langle
	\hat{\Psi}^{\dagger}(x',t)\hat{\Psi}(x,t)
	\rangle.
\end{equation}
The natural orbitals $\phi_j(x,t)$ and their corresponding occupations $\lambda_j(t)$ are obtained from the eigenvalue equation
\begin{equation}
	\int \rho^{(1)}(x,x';t)\phi_j(x',t)\;\mathrm{d}x'
	=
	\lambda_j(t)\phi_j(x,t).
\end{equation}
According to the Penrose--Onsager criterion, the condensate is identified with the natural orbital corresponding to the largest eigenvalue $\lambda_0(t)$~\cite{penrose1956bose,pethick2002bose}. The condensate fraction is therefore given by
\begin{equation}
	\bar{\lambda}(t)=\frac{\lambda_0(t)}{\mathcal{N}},
\end{equation}
where $\mathcal{N}$ is the total number of atoms. A reduction in $\bar{\lambda}(t)$ indicates depletion of the condensate into the non-condensed modes. Within TWA, the field operator is represented by an ensemble of stochastic classical fields $\psi_{\mathcal{C},i}(x,t)$, with the appropriate Wigner-vacuum correction, and subsequently averaged over the ensemble~\cite{blakie2008dynamics}. 

The evolution of the condensate fraction for different nonlinearities and atom numbers is shown in Fig~(\ref{cond_fr}). At weak nonlinearity, the condensate fraction remains close to unity throughout the evolution. As the nonlinearity is increased, the condensate fraction decreases, indicating an increasing transfer of population from the condensate mode to the non-condensed modes. This reduction is more pronounced for smaller atom numbers at a fixed value of $\mathrm{g}\mathcal{N}$, where the relative contribution of quantum fluctuations is larger. Interestingly, the condensate fraction exhibits partial revivals at the FR instances, indicating a partial recovery of the macroscopic occupation of the condensate mode. These partial revivals become more pronounced at higher nonlinearities [Fig.~\ref{cond_fr}(c)-(e)], particularly for lower atom numbers.

The condensate fraction quantifies the population remaining in the most highly occupied condensate mode, but does not characterize the phase alignment between different stochastic realizations. To characterize ensemble dephasing, we consider the overlap of each TWA realization with the mean-field condensate mode,
\begin{equation}
	a_i(t)=
	\int
	\psi_{0}^{*}(x,t)\psi_{\mathcal{C},i}(x,t)\;\mathrm{d}x.
\end{equation}
The complex coefficient $a_i(t)$ contains the amplitude and phase of the $i$th TWA realization projected onto the mean-field condensate mode. The corresponding phase coherence is defined as
\begin{equation}
	c(t)=
	\frac{\left|\left\langle a_i(t)\right\rangle\right|^2
	}{\left\langle |a_i(t)|^2\right\rangle}.
\end{equation}

For an ensemble in which the TWA realizations remain phase aligned, the complex amplitudes $a_i(t)$ have approximately the same phase and $c(t)$ remains close to unity. As the realizations acquire different phases during the nonlinear evolution, the distribution of $a_i(t)$ broadens in phase and the ensemble-averaged amplitude decreases, resulting in a reduction of $c(t)$. Thus, $c(t)$ provides a measure of the phase coherence of the TWA ensemble.

The time evolution of $c(t)$ for different nonlinearities is depicted in Fig.~(\ref{phase_coh}). For large atom numbers, $\mathcal{N}=10000,\;15000,\;20000$, the phase coherence remains close to unity for nonlinearities up to $\mathrm{g}\mathcal{N}=20$, indicating that the TWA ensemble remains largely phase coherent with the mean-field evolution. However, for $\mathrm{g}\mathcal{N}>30$, a substantial reduction in the phase coherence is observed, with $c(t)$ decreasing to approximately $0.7$. In contrast, for smaller atom numbers, $\mathcal{N}=1000,\;1500,\;2000$, there is a substantial reduction in $c(t)$ even at $\mathrm{g}\mathcal{N}=10$. This behavior reflects the larger relative contribution of the quantum fluctuations for smaller condensate populations. As the nonlinearity is increased, $c(t)$ decreases progressively with time due to the accumulation of different nonlinear phases across the ensemble. At sufficiently strong nonlinearity ($\mathrm{g}\mathcal{N}>30$), the phase coherence is strongly suppressed before $t=T_{\mathrm r}/4$, indicating substantial ensemble dephasing within the timescale required for the quarter revival.

\begin{figure}[ht!]
	\centering
	\includegraphics[width=15 cm]{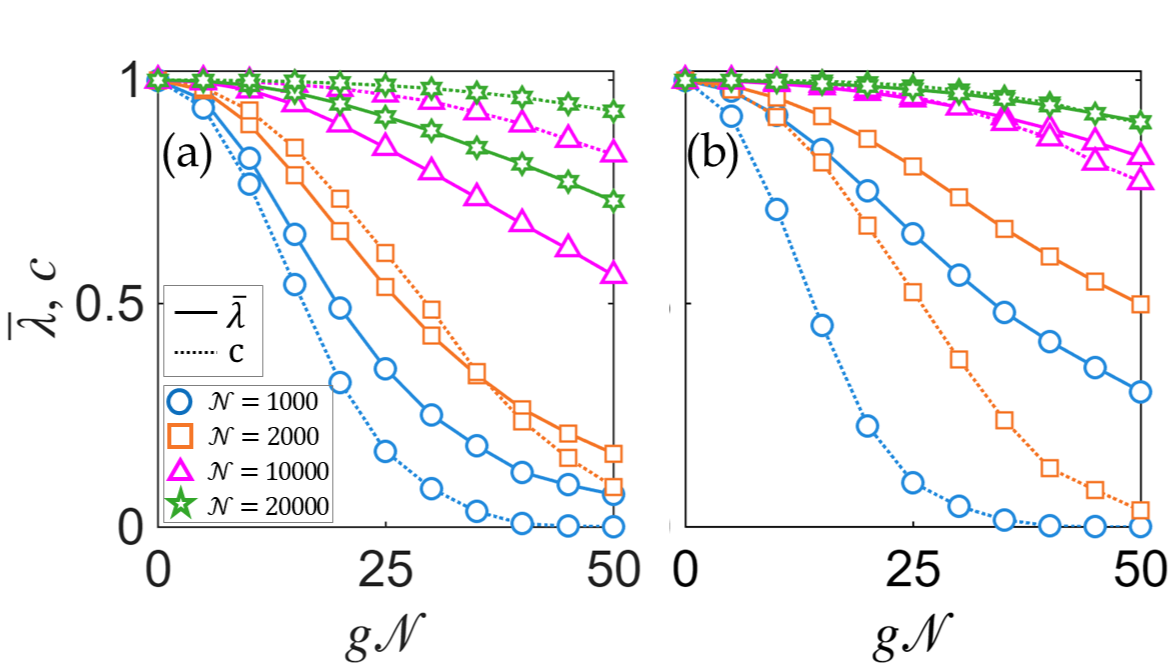}
	\caption{Comparison of the condensate fraction $\bar{\lambda}$ and phase coherence $c$ as a function of $\mathrm{g}\mathcal{N}$ at (a) $t=0.425T_r$ and at (b) $t=0.5T_r$ for both small atom numbers $\mathcal{N}=1000,\;2000$ and large atom numbers $\mathcal{N}=10000,\;20000$. The condensate fraction $\bar{\lambda}$ is represented by solid lines, while the phase coherence $c$ is represented by dotted lines. Time is expressed in units of $\omega_x^{-1}=3.18\,\mathrm{ms}$.}
	\label{c_lam_comb}
\end{figure}

The combined dependence of $\bar{\lambda}$ and $c$ on the interaction strength is shown in Fig.~(\ref{c_lam_comb}) at selected fractional-revival times: $t=0.425T_r$ in (a), and $t=0.5T_r$ in (b). At $t=0.425T_r$, for $\mathcal{N}=2000$, the degradation of $c$ and $\bar{\lambda}$ is comparable. For $\mathcal{N}<2000$, $c$ decays faster than $\bar{\lambda}$, whereas for higher atom numbers $\mathcal{N}=10000$ and $20000$, the condensate fraction decreases more strongly than the phase coherence.

At $t=0.5T_r$, a clearer separation between the behavior of $c$ and $\bar{\lambda}$ is observed. For low atom numbers, the phase coherence decreases to nearly zero despite a substantial condensate fraction, indicating that a large condensate population does not necessarily imply phase coherence across the TWA ensemble. Interestingly, although the condensate fraction exhibits partial revivals at the FR instances, no corresponding recovery of the phase coherence is observed. This indicates that the fractional-revival dynamics can partially restore the macroscopic occupation of the condensate mode without reversing the accumulated phase dispersion across the TWA ensemble. In contrast, for large atom numbers, both $c$ and $\bar{\lambda}$ remain close to unity, indicating that both ensemble dephasing and condensate depletion are strongly suppressed.

\subsection{FR Interferometry}\label{sss4.1_c7}

From the analysis of phase coherence with respect to the condensate mode, we find that the phase misalignment across the TWA ensemble increases with time and with the nonlinearity. The condensate fraction also decreases with increasing $\mathrm{g}\mathcal{N}$. These effects can degrade the coherence of the FR interferometer and therefore motivate the consideration of small to moderate nonlinearities. As discussed earlier, in FR interferometry, an initially localized BEC undergoes a coherent $50:50$ splitting at the quarter revival, followed by recombination at the half revival. Since the revival time decreases with increasing $\mathrm{g}\mathcal{N}$, the corresponding times for the $50:50$ splitting and recombination are also modified in the presence of interactions. We therefore determine the quarter-revival time $t_{\mathrm{qr}}$ numerically for each value of $\mathrm{g}\mathcal{N}$.

Here, we analyze the phase statistics of the two spatially separated condensates at the quarter revival as a function of the interaction strength. The mode amplitudes of the two clouds are defined as \cite{isella2006quantum,martin2012quantum}
\begin{eqnarray}
	a_{i}&=&\int_{-\pi R}^{\pi R}\text{d}x\;
	\psi^{*}_{a}(x,t_{\mathrm{qr}})
	\psi^{(i)}_{\mathrm{TW}}(x,t_{\mathrm{qr}})\nonumber\\
	b_{i}&=&\int_{-\pi R}^{\pi R}\text{d}x\;
	\psi^{*}_{b}(x,t_{\mathrm{qr}})
	\psi^{(i)}_{\mathrm{TW}}(x,t_{\mathrm{qr}}) \label{mode_amps}
\end{eqnarray}
where
\begin{equation}
	\psi_{a,b}(x,t_{\mathrm{qr}})
	=
	\frac{\psi_{\mathrm{GP}}(x,t_{\mathrm{qr}})\Xi_{a,b}(x)}
	{\sqrt{
			\int\limits_{-\pi R}^{\pi R}\text{d}x\;
			\left|\psi_{\mathrm{GP}}(x,t_{\mathrm{qr}})
			\Xi_{a,b}(x)\right|^2
	}}.\label{window}
\end{equation}
Here, $\Xi_{a,b}(x)$ are the spatial window functions corresponding to $|x|\leq\pi R/2$ and $|x|\geq\pi R/2$, respectively. The time $t_{\mathrm{qr}}$ denotes the quarter-revival time, which depends on the interaction strength $\mathrm{g}\mathcal{N}$ as discussed in Fig.~\ref{AC}. Around the quarter revival, the two clouds are spatially separated and the two-mode approximation is therefore applicable, with
\begin{equation}
	\int\limits_{-\pi R}^{\pi R}\text{d}x\;
	\psi^{*}_{a}(x,t_{\mathrm{qr}})
	\psi_{b}(x,t_{\mathrm{qr}})
	=0.
\end{equation}
The same procedure can also be applied at other FR instances at which two spatially separated clouds are obtained, such as $t=pT_{\mathrm r}/4$, where $p$ is an integer.

From these mode amplitudes, we characterize the coherence between the two spatially separated clouds using the normalized first-order correlation function and the relative-phase coherence \cite{isella2006quantum},
\begin{equation}\label{relative_phasecoh}
	c_{12}
	=
	\frac{
		\left|\left\langle a^{*}_{i}b_{i}\right\rangle
		\right|
	}{
		\sqrt{\mathcal{N}_{1}\mathcal{N}_{2}}
	},
	\qquad
	\left|\left\langle e^{i\phi_{12}}\right\rangle\right|
	=
	\left|
	\left\langle
	e^{i[\arg(a_{i})-\arg(b_{i})]}
	\right\rangle
	\right|.
\end{equation}
Here, $\mathcal{N}_{1}=\langle|a_{i}|^2\rangle-1/2$ and $\mathcal{N}_{2}=\langle|b_{i}|^2\rangle-1/2$ are the mean populations of the two modes. The first-order correlation satisfies $c_{12}=1$ for perfectly coherent modes and approaches zero as first-order coherence is lost. The relative-phase coherence, $\left|\langle e^{i\phi_{12}}\rangle\right|$, characterizes the phase locking between the two modes. It is unity when the relative phase is well defined and approaches zero when the relative phase becomes randomized. In general, these two quantities need not have the same value, since the first-order correlation depends on both the amplitude and phase fluctuations of the two modes, whereas $\left|\langle e^{i\phi_{12}}\rangle\right|$ depends only on their relative phase.

\begin{figure}[ht!]
	\centering
	\includegraphics[width=15 cm]{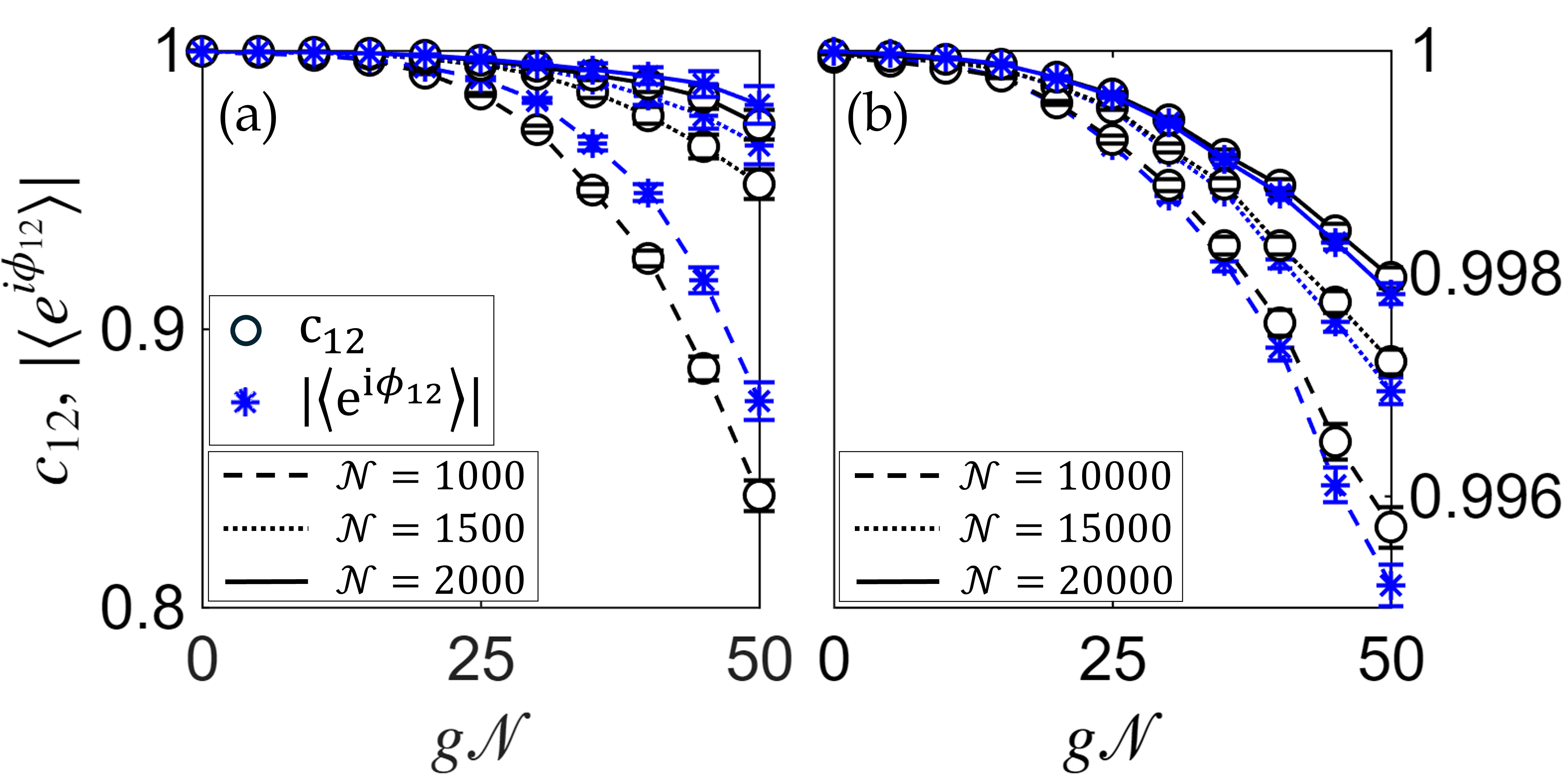}
	\caption{Comparison of the first-order correlation function $c_{12}$ and relative-phase coherence $|\langle e^{\iota\phi_{12}}\rangle|$ as a function of $\mathrm{g}\mathcal{N}$ for (a) small atom numbers $\mathcal{N}=1000,\;1500,\;2000$ and for (b) large atom numbers $\mathcal{N}=10000,\;15000,\;20000$. The $1^{\mathrm{st}}$-order correlation function is represented by circles, while the relative-phase coherence is denoted by asterisk. Time is expressed in units of $\omega_x^{-1}=3.18\,\mathrm{ms}$.}
	\label{phase_c12}
\end{figure}

Figure~\ref{phase_c12} shows the variation of $c_{12}$ and the relative-phase coherence with the interaction strength $\mathrm{g}\mathcal{N}$ for small and large atom numbers. For the smallest atom number, $\mathcal{N}=1000$, both $c_{12}$ and the relative-phase coherence decrease appreciably with increasing nonlinearity, reaching values of approximately $0.85$ at the largest $\mathrm{g}\mathcal{N}$ considered. For $\mathcal{N}=1500$ and $2000$, the reduction is less pronounced, with both quantities remaining above approximately $0.95$. In contrast, for the larger atom numbers $\mathcal{N}=10000,\;15000,\;20000$ shown in Fig.~\ref{phase_c12}(b), both the first-order correlation and relative-phase coherence remain very close to unity over the entire range of $\mathrm{g}\mathcal{N}$. Thus, the coherence between the two FR-generated clouds is strongly dependent on the atom number, with smaller condensates being more susceptible to interaction-induced fluctuations.

\subsubsection{Possibility of sub-shot noise interferometry}

Having established that the FR dynamics maintains a highly coherent $50:50$ splitting for moderate to large atom numbers, we now investigate the possibility of realizing a sub-shot-noise nonlinear FR interferometer. Sub-shot-noise atom interferometry can provide enhanced sensitivity over the shot-noise limit by exploiting quantum correlations between the two interferometer modes~\cite{Wineland1992,pezze2018quantum}.

At the quarter revival, the two spatially separated clouds can be described as two modes, $\psi_{a}(x,t_{\mathrm{qr}})$ and $\psi_{b}(x,t_{\mathrm{qr}})$, containing a total of $\mathcal{N}$ atoms. The two-mode system can therefore be represented by a collective spin of length $J=\mathcal{N}/2$~\cite{schwinger1952angular,gross2012spin}. The collective spin operators $\hat{J}_{\mu}$, with $\mu=\{x,y,z\}$, satisfy the angular momentum commutation relations
\begin{equation}
	[\hat{J}_{\mu},\hat{J}_{\nu}]
	=
	i\epsilon_{\mu\nu\rho}\hat{J}_{\rho},
\end{equation}
where $\epsilon_{\mu\nu\rho}$ is the Levi-Civita symbol. In the Schwinger representation, these operators are written as~\cite{gross2012spin,haine2018quantum}
\begin{eqnarray}
	\hat{J}_x&=&\frac{1}{2}
	(\hat{a}^{\dagger}\hat{b}+\hat{a}\hat{b}^{\dagger})\nonumber\\
	\hat{J}_y&=&\frac{1}{2i}
	(\hat{a}^{\dagger}\hat{b}-\hat{a}\hat{b}^{\dagger})\nonumber\\
	\hat{J}_z&=&\frac{1}{2}
	(\hat{a}^{\dagger}\hat{a}-\hat{b}^{\dagger}\hat{b}).
\end{eqnarray}
Here, the annihilation operators $\hat{a}$ and $\hat{b}$ correspond to atoms occupying the two modes $\psi_{a}(x,t_{\mathrm{qr}})$ and $\psi_{b}(x,t_{\mathrm{qr}})$, respectively. Within the TWA, the corresponding operators are represented by the mode amplitudes obtained from Eq.~\eqref{mode_amps}. The two-mode state can therefore be represented by a collective spin on the generalized Bloch sphere with spin length $J=\mathcal{N}/2$.

\begin{figure}[ht!]
	\centering
	\includegraphics[width=15 cm]{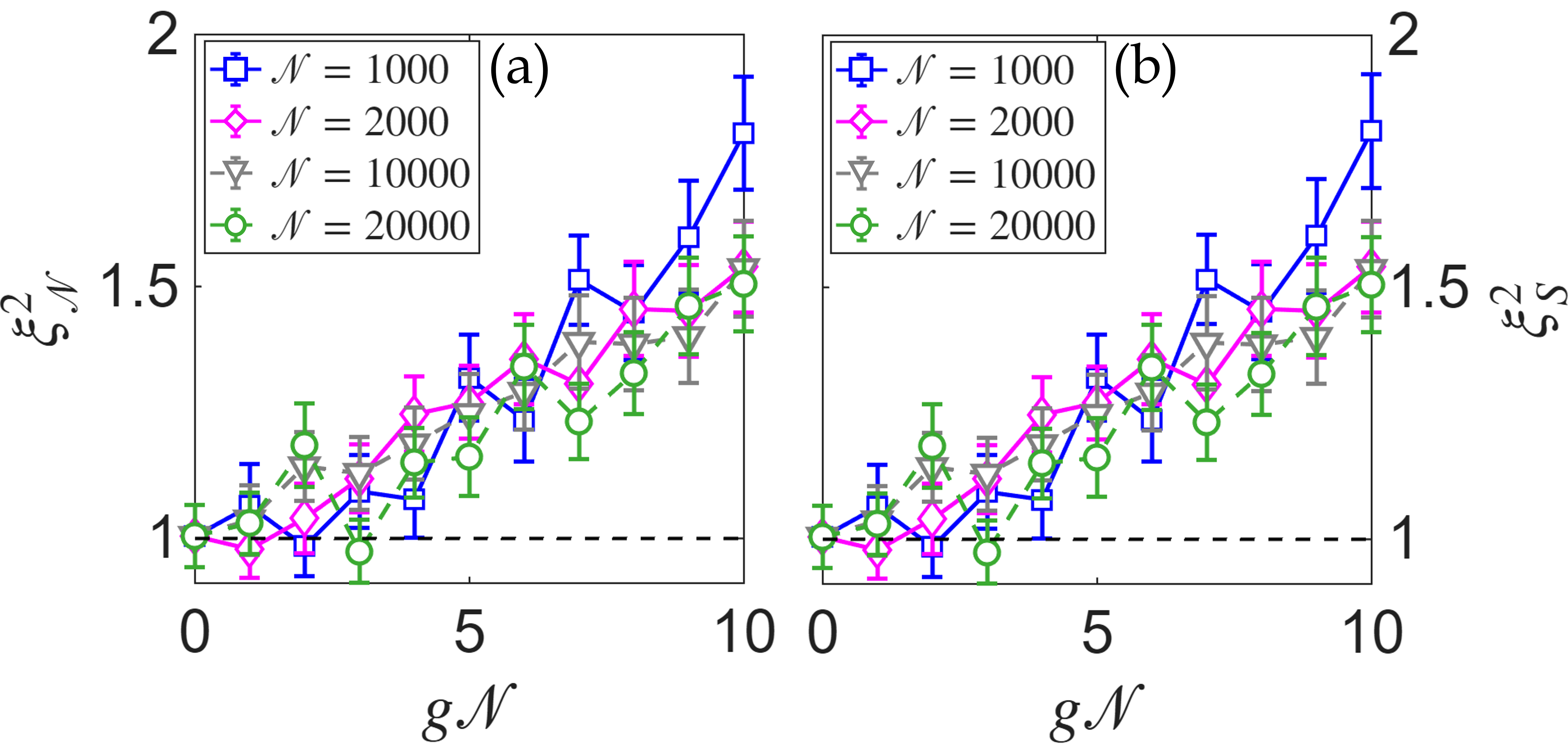}
	\caption{Variation of (a) number-squeezing parameter $\xi^2_{\mathcal{N}}$ and (b) spin-squeezing parameter $\xi^2_{\mathcal{S}}$ with respect to $\mathrm{g}\mathcal{N}$ for different atom numbers $\mathcal{N}$. The dashed-line indicates the standard quantum limit in both the cases.}
	\label{num_sq}
\end{figure}
The relative-number fluctuations between the two split BECs at the quarter revival are characterized by the number-squeezing parameter~\cite{martin2012quantum}
\begin{equation}
	\xi^2_{\mathcal{N}}
	=
	\mathrm{var}(\hat{\delta\mathcal{N}})\frac{
		\langle
		\hat{\mathcal{N}}_1+\hat{\mathcal{N}}_2
		\rangle}{4\langle\hat{\mathcal{N}}_1\rangle\langle\hat{\mathcal{N}}_2\rangle},
\end{equation}
where
\begin{equation}
	\hat{\delta\mathcal{N}}
	=
	\hat{\mathcal{N}}_2-\hat{\mathcal{N}}_1.
\end{equation}
For a balanced splitting with binomial number fluctuations, $\xi^2_{\mathcal{N}}=1$ defines the shot-noise limit. Interactions can modify the relative-number fluctuations, producing number-squeezed states with $\xi^2_{\mathcal{N}}<1$ or number anti-squeezed states with $\xi^2_{\mathcal{N}}>1$.

The number-squeezing parameter alone does not account for the phase coherence between the two split BECs. Therefore, reduced relative-number fluctuations do not necessarily result in useful interferometric squeezing if the phase coherence between the two modes is strongly degraded. To incorporate the phase coherence through the transverse mean spin, we use the spin-squeezing parameter~\cite{martin2012quantum}
\begin{equation}
	\xi^2_{\mathcal{S}}
	=
	\frac{
		\mathcal{N}\Delta\hat{J}^{2}_{z}
	}{
		\langle\hat{J}_{x}\rangle^{2}
		+
		\langle\hat{J}_{y}\rangle^{2}
	}.
\end{equation}
This parameter compares the relative-number fluctuations, represented by $\Delta\hat{J}_{z}^{2}$, with the squared transverse mean spin and therefore accounts for the reduction in interferometric contrast arising from imperfect phase coherence. For highly phase coherent splitting, $\xi^2_{\mathcal{N}}\approx\xi^2_{\mathcal{S}}$.

Figure~\ref{num_sq}(a) shows the dependence of $\xi^2_{\mathcal{N}}$ on $\mathrm{g}\mathcal{N}$ at the quarter-revival time for both small and large atom numbers. In the non-interacting limit, $\xi^2_{\mathcal{N}}\approx1$, corresponding to the shot-noise limit. With increasing nonlinearity, $\xi^2_{\mathcal{N}}$ increases above unity, indicating that the nonlinear evolution does not generate useful number squeezing; instead, the relative-number fluctuations become anti-squeezed. The overall increase is similar for the different atom numbers considered. Figure~\ref{num_sq}(b) shows the corresponding behavior of $\xi^2_{\mathcal{S}}$. Its dependence on $\mathrm{g}\mathcal{N}$ closely follows that of $\xi^2_{\mathcal{N}}$, consistent with the high phase coherence of the two split condensates over the parameter range considered.

\subsubsection{Minimum Noise Axis and Wineland Spin Squeezing}

The squeezing parameters $\xi^2_{\mathcal{N}}$ and $\xi^2_{\mathcal{S}}$ discussed above quantify variance along a fixed laboratory axis (typically $\hat{J}_z$). However, during the nonlinear evolution in a BEC, the quantum noise ellipse in the transverse plane rotates away from the Cartesian axes~\cite{kitagawa1993squeezed}. For the present choice of the two modes at the quarter revival, the mean spin lies within the $x\text{--}y$ plane with $\langle\hat{J}_z\rangle \approx 0$, so that the $J_y\text{--}J_z$ plane is strictly the transverse plane. To determine the optimal squeezing direction, we construct the transverse covariance matrix~\cite{gross2012spin}:
\begin{equation}
	\Gamma_{yz}
	=
	\begin{pmatrix}
		\Delta\hat{J}^{2}_{y} & \mathrm{Cov}(\hat{J}_{y},\hat{J}_{z}) \\
		\mathrm{Cov}(\hat{J}_{y},\hat{J}_{z}) & \Delta\hat{J}^{2}_{z}
	\end{pmatrix},
\end{equation}
with the covariance defined via the symmetrized operator product:
\begin{equation}
	\mathrm{Cov}(\hat{J}_{y},\hat{J}_{z}) = \frac{1}{2}\left\langle \{ \hat{J}_{y}, \hat{J}_{z} \} \right\rangle - \langle\hat{J}_{y}\rangle\langle\hat{J}_{z}\rangle.
\end{equation}

Diagonalizing $\Gamma_{yz}$ yields the variances along the principal directions of the noise ellipse~\cite{kitagawa1993squeezed}. The minimum variance $\Delta \hat{J}_{\min}^{2}$ corresponds to the smaller eigenvalue of $\Gamma_{yz}$:
\begin{equation}
	\Delta \hat{J}_{\min}^{2} = \frac{\Delta\hat{J}_{y}^{2} + \Delta\hat{J}_{z}^{2}}{2} - \sqrt{\left(\frac{\Delta\hat{J}_{y}^{2} - \Delta\hat{J}_{z}^{2}}{2}\right)^{2} + \mathrm{Cov}(\hat{J}_{y},\hat{J}_{z})^{2}}.
\end{equation}
Consider a quadrature axis rotated by an angle $\alpha$ relative to the $J_y$-axis, $\hat{J}_{\alpha} = \hat{J}_{y}\cos\alpha + \hat{J}_{z}\sin\alpha$. The extrema of $\Delta\hat{J}_{\alpha}^{2}$ determine the two principal
axes of the noise ellipse and satisfy
\begin{equation}
	\tan(2\alpha) = \frac{2\,\mathrm{Cov}(\hat{J}_{y},\hat{J}_{z})}{\Delta\hat{J}_{y}^{2} - \Delta\hat{J}_{z}^{2}}.
\end{equation}
The branch corresponding to the smaller eigenvalue defines the principal squeezing angle $\alpha_{\mathrm{min}}$. In our numerical simulation, $\alpha_{\mathrm{min}}$ is extracted directly from the eigenvector associated with $\Delta \hat{J}_{\min}^{2}$ and is defined modulo $\pi$.

To evaluate the metrological gain achieved along this optimal direction, we introduce the Wineland spin-squeezing parameter $\xi_R^2$~\cite{Wineland1992,pezze2018quantum}:
\begin{equation}
	\xi_{R}^{2} = \frac{N\,\Delta\hat{J}_{\min}^{2}}{|\langle\hat{\mathbf{J}}\rangle|^{2}},
\end{equation}
where $|\langle\hat{\mathbf{J}}\rangle|^2 = \langle\hat{J}_{x}\rangle^{2} + \langle\hat{J}_{y}\rangle^{2} + \langle\hat{J}_{z}\rangle^{2}$ is the total length of the mean spin vector. Since $\langle\hat{J}_{z}\rangle\simeq0$ at the quarter revival, the
mean-spin length is, to the same accuracy, $|\langle\hat{\mathbf{J}}\rangle|^2 \simeq \langle\hat{J}_{x}\rangle^{2} + \langle\hat{J}_{y}\rangle^{2}$. Under this condition, the Wineland parameter is identical to the optimized transverse parameter $\xi_{\min}^{2}$, satisfying:
\begin{equation}
	\xi_R^{2} \simeq \xi_{\min}^{2} = \frac{N\,\Delta\hat{J}_{\min}^{2}}{\langle\hat{J}_{x}\rangle^{2} + \langle\hat{J}_{y}\rangle^{2}} \le \xi_{\mathcal{S}}^{2}.
\end{equation}
Hence, allowing the squeezing axis to rotate to the minimum-noise direction can only improve upon, or recover, the squeezing obtained when the population-difference axis $\hat{J}_z$ is used directly. 
\begin{figure*}[ht!]
	\centering
	\includegraphics[width=15 cm]{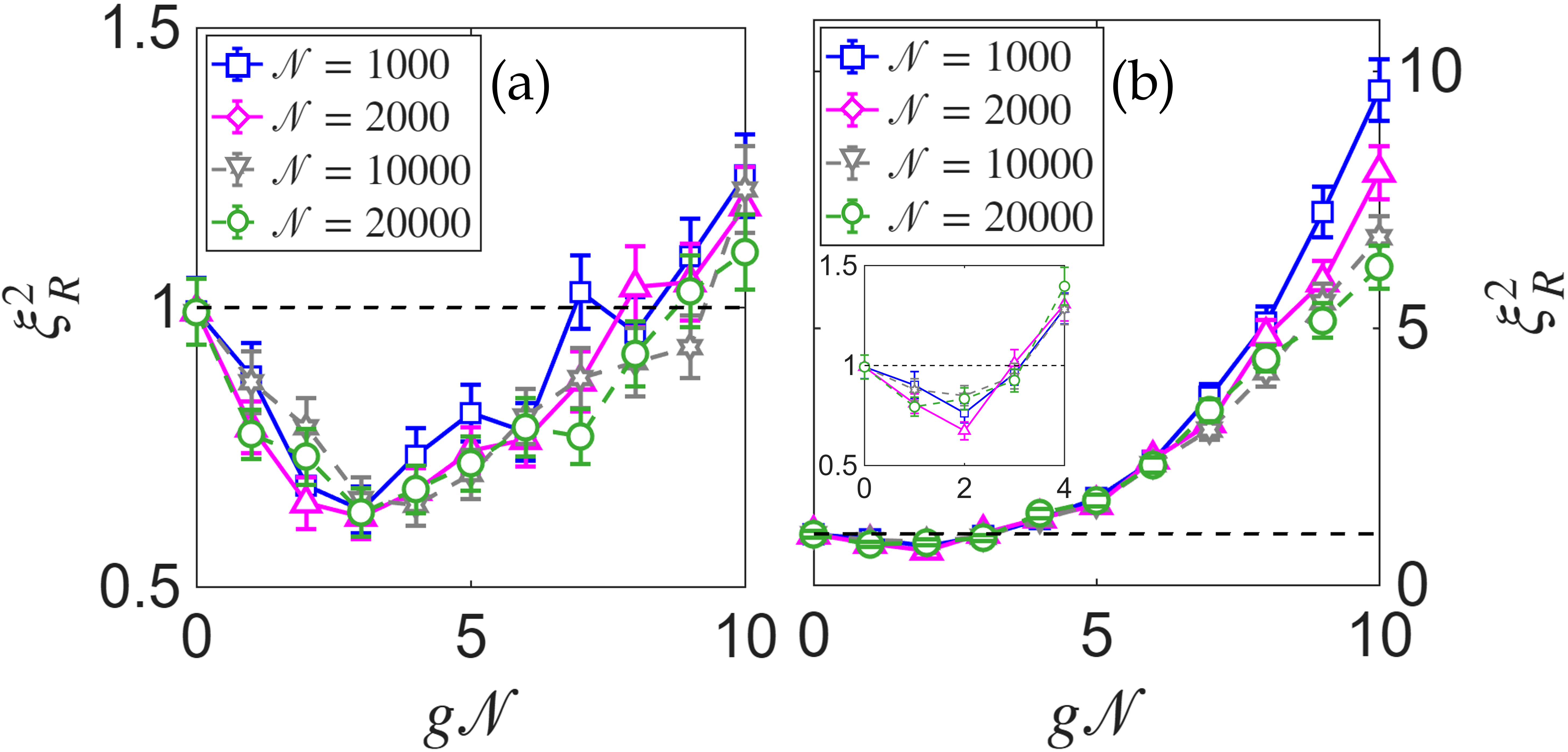}
	\caption{Wineland squeezing paramter as a function of nonlinearity $\mathrm{g}\mathcal{N}$ for various atom numbers at (a) $t=0.25T_{r}$ and (b) $t=0.5T_{r}$. The inset shows the zoomed in portion near the nonlinearity range $\mathrm{g}\mathcal{N}=0$ to $4$. The dashed-line indicates the standard quantum limit.}
	\label{wineland_NL}
\end{figure*}

In order to determine if the nonlinear evolution of the BEC to
the quarter revival generates appreciable spin squeezing along an
optimally chosen transverse direction, we calculate the Wineland
squeezing parameter $\xi_R^2$ at $t=0.25T_r$. In Fig.~\ref{wineland_NL}, we have shown the variation of $\xi^2_{R}$ with respect to the nonlinearity $\mathrm{g}\mathcal{N}$. The black dashed line shows the shot-noise limit corresponding to $\xi_{R}^2=1$. For all atom numbers considered, the minimum occurs around $\mathrm{g}\mathcal{N}=3$, with $\xi_R^2$ remaining close to $0.63$. For $\mathcal{N}=2000$, we obtain a minimum value of $\xi_{R}^2=0.626\pm0.04$, corresponding to a squeezing of $-2.03\pm0.28\;$dB. The principal squeezing angle $\alpha_{\mathrm{min}}$ corresponding to this value is found to be $126.42^\circ\pm2.08^\circ$. However, as we increase nonlinearity, $\xi^2_{R}$ increases and crosses the shot-noise limit even at around $\mathrm{g}\mathcal{N}=8$. We observe that the behavior is almost identical for both small and large atom numbers.

The spin squeezing obtained at the quarter revival may not be described by
the usual two-mode One-Axis Twisting (OAT) picture. In conventional two-mode OAT, one starts
from a coherent spin state and turns on the nonlinear interaction,
which shears the spin distribution in the transverse plane and
generates spin squeezing \cite{kitagawa1993squeezed}. In the present protocol, however, the
initial state is already an interacting BEC, and the same nonlinear
interaction remains present throughout the evolution leading to the
quarter revival. Moreover, the two-mode description is valid only at
the fractional revival times $t=pT_r/4$, where the dynamics can be
projected onto the two spatially separated modes. At intermediate
times, the wavefunction occupies multiple spatial modes and cannot be
faithfully described by a two-mode model. Consequently, the spin
squeezing observed at the quarter revival emerges from the full
multimode nonlinear dynamics rather than from conventional two-mode
OAT.

The above results obtained through nonlinear evolution up to the
quarter revival can be useful for sub-shot-noise interferometry if
the interactions responsible for generating the squeezing is turned
off after $t=0.25T_r$. Turning off the interactions at this point
freezes the spin ellipse with its minimum-noise axis oriented at an
angle $\alpha_{\mathrm{min}}$. For instance, for
$\mathrm{g}\mathcal{N}=3$ and $\mathcal{N}=2000$, shown in
Fig.~\ref{wineland_NL}(a), turning off the interactions at
$t=0.25T_r$ freezes the minimum-noise axis at
$\alpha_{\mathrm{min}}=126.42^\circ\pm2.08^\circ$. A subsequent
$-\alpha_{\mathrm{min}}$ rotation of the spin ellipse about the
$J_x$ axis aligns the minimum-noise quadrature with the phase-sensitive
direction, resulting in a phase-squeezed state in which an unknown
phase can be stored with minimum variance. Further, another quarter
revival evolution converts the phase-squeezed state into a
number-squeezed state, allowing the phase information to be read out
from the population imbalance. The exact protocol for implementing
the $\alpha_{\mathrm{min}}$ rotation about the $J_x$ axis is discussed
in the next section. Given such a rotation, a spin-squeezed state with
Wineland parameter $\xi_R^2=0.63$ corresponds to a phase sensitivity
of
\[
\Delta\Phi
=
\frac{\xi_R}{\sqrt{\mathcal{N}}}
=
0.79\,\Delta\Phi_{\mathrm{SQL}},
\]
where
$\Delta\Phi_{\mathrm{SQL}}=1/\sqrt{\mathcal{N}}$ denotes the
standard quantum limit.

In the above protocol, turning off the nonlinear interactions at $t=0.25T_r$ is crucial, as further evolution in the presence of nonlinearity not only reduces the condensate fraction and phase coherence, as discussed in Figs.~\ref{cond_fr} and \ref{phase_coh}, but also degrades the optimal spin squeezing. This is clear from Fig.~\ref{wineland_NL}(b), where the Wineland squeezing parameter as a function of $\mathrm{g}\mathcal{N}$ is shown at $t=0.5T_r$, after imprinting a phase of $\pi/2$ on one of the clouds at $t=0.25T_r$. The crossover to $\xi_R^2=1$ now occurs at an even lower nonlinearity, around $\mathrm{g}\mathcal{N}=3$, compared with $\mathrm{g}\mathcal{N}=8$ at the quarter revival. The maximum squeezing occurs at $\mathrm{g}\mathcal{N}=2$ for $\mathcal{N}=2000$, with a Wineland parameter $\xi_R^2=0.674\pm0.045$. Here, the minimum-noise axis is aligned at an angle $\alpha_{\mathrm{min}}=174^\circ\pm2.4^\circ$ with respect to the $J_y$ axis. 

\section{Interaction-quenched FR-Interferometry for enhanced phase sensitivity}

The preceding section showed that continuous nonlinear evolution to the quarter revival yields only a modest maximum squeezing of $-2\;$ dB, further eroded by condensate depletion and ensemble dephasing (Figs.~\ref{cond_fr}–\ref{wineland_NL}). This raises an important question: \textit{how much nonlinear evolution is actually required to prepare a useful spin-squeezed state?}

Unlike the soliton-based interferometer, where the attractive
interaction is required throughout the interferometric cycle~\cite{martin2012quantum,haine2018quantum}, the
present FR-based protocol requires the interaction only during the
spin-squeezing preparation stage. The subsequent interferometric
operations, including the generalized beam-splitter operations, can be
performed using FR dynamics. We therefore consider an
interaction-quenched protocol in which a short nonlinear evolution is
used to generate spin squeezing, followed by a quench of the
interaction. The remaining interferometric operations are then
performed using FR dynamics without nonlinearity.

In conventional two-mode one-axis twisting (OAT), the nonlinear
interaction is described by the effective Hamiltonian~\cite{kitagawa1993squeezed}
\begin{equation}
	\hat{H}_{\mathrm{OAT}}
	=
	\hbar\chi\hat{J}_{z}^{2},
\end{equation}
where $\chi$ characterizes the nonlinear interaction strength and $\hat{J}_{z}=\frac{1}{2}(\hat{N}_{1}-\hat{N}_{2})$ is the population-imbalance operator. The $\hat{J}_{z}^{2}$ interaction
introduces a nonlinear phase that depends on the population imbalance,
shearing an initially coherent spin distribution in the transverse
plane and thereby generating spin squeezing.

In the present protocol, we realize an OAT-like nonlinear squeezing
stage directly within the FR dynamics. At the quarter revival,
$t=T_r/4$, the two spatially separated modes are subjected to the
nonlinear interaction for a short duration $\Delta t$. The duration
$\Delta t$ is chosen sufficiently small that the two modes remain
spatially separated and significant mode mixing does not occur. The
interaction is then quenched at
$t=T_r/4+\Delta t$, thereby preserving the spin squeezing generated
during this nonlinear stage. Importantly, the squeezing dynamics
reported below are obtained directly from the TWA simulations;
the OAT picture is used only to provide a physical interpretation of
the short nonlinear evolution.

\begin{figure*}[ht!]
	\centering
	\includegraphics[width=16 cm]{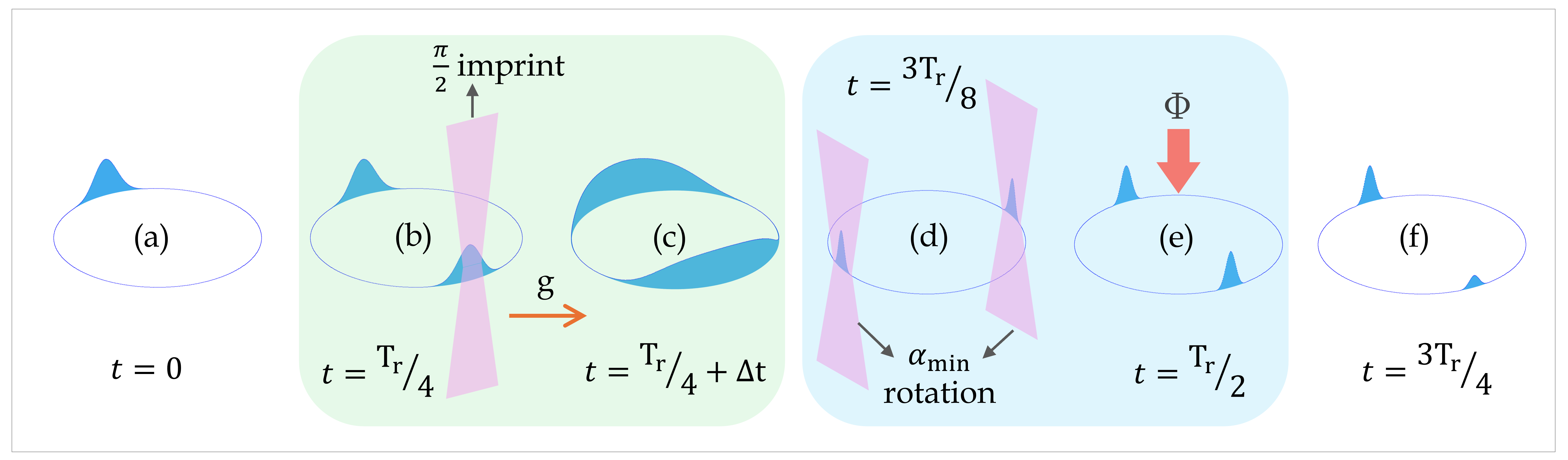}
	\caption{Quantum enhanced FR interferometry protocol. (a) Initial non-interacting cloud of BEC, (b) $50:50$ splitting at quarter revival $t=0.25T_r$ with a $\pi/2$ phase imprint on one of the clouds  (c) Switching on the interactions for a duration $\Delta t$ which shears the noise distribution with minimum noise axis aligned at $\alpha_{\mathrm{min}}$ with the $J_y$ axis. (d) A phase imprint of $\alpha_{\mathrm{min}}-\pi/2$ at
		$t=3T_r/8$, followed by a quarter revival, rotates the minimum-noise
		quadrature into the phase-sensitive direction and produces a
		phase-squeezed state at $t=T_r/2$, where the unknown
		interferometric phase $\Phi$ is encoded.(f) A subsequent quarter revival to $t=3T_r/4$ gives us the number-squeezed state from which the interferometric phase is read-out as population difference. Here,the green shaded portion highlights the part where spin-squeezing is generated and the blue shaded region illustrates the rotation of the optimal spin-squeezing towards the $J_{y}$ axis.} 
	\label{sch_SS}
\end{figure*}
\subsection{Protocol}
The complete interferometric sequence is illustrated in
Fig.~\ref{sch_SS}. We begin with a non-interacting BEC [Fig.~\ref{sch_SS}(a)] and use the quarter-revival dynamics to generate two equally populated spatially separated clouds at $t=T_r/4$.  This state corresponds to a coherent spin state, whose noise distribution is approximately circular in the transverse $J_y-J_z$ plane [Fig.~\ref{sch_SS}(b)]. The nonlinear interaction is then applied for a duration $\Delta t$, resulting in the shearing of the spin distribution and the generation of a minimum-noise quadrature oriented at $\alpha_{\mathrm{min}}$ with respect to $J_y$ axis [Fig.~\ref{sch_SS}(c)].
After the interaction is quenched, a phase imprint of
$\alpha_{\mathrm{min}}-\pi/2$ at $t=3T_r/8$, followed by a quarter revival, rotates the squeezed quadrature into the phase-sensitive direction, producing a phase-squeezed state at $t=T_r/2$ [Fig.~\ref{sch_SS}(d,e)]. An unknown interferometric phase $\Phi$ is then encoded in this state. A further quarter revival converts the phase-squeezed state into a number-squeezed state at $t=3T_r/4$, from which the interferometric phase can be read out through the population imbalance [Fig.~\ref{sch_SS}(f)].

The connection between the physical phase of the two clouds and the collective-spin representation requires some care because of our definition of the mode amplitudes in Eq.\eqref{mode_amps}. We define the mode amplitudes
$a_{i}$ and $b_{i}$ such that the relative phase between the two spatially separated clouds is absorbed into the mode amplitudes. Consequently, for a balanced splitting, the mean spin is represented along the $J_x$ axis and the $J_y$--$J_z$ plane is the transverse
plane. Physically, the relative phase between the two clouds at the quarter revival is $\pi/2$, as shown in
Fig.~\ref{FR_phase_den_NI}(b), which would conventionally correspond to a mean spin oriented along $J_y$. In our mode convention, however, this phase is absorbed into $a_{i}$ and $b_{i}$, and the same state is represented with the mean spin along $J_x$. This convention is used
consistently throughout the definition of the collective-spin
operators.

In terms of the unitary operators introduced in Sec.~\ref{UnitOp}, the spin-squeezing preparation stage shown in Fig.~\ref{sch_SS}(a-c) can be written as
\begin{equation}
	U_{\mathrm{SS}}
	=
	U_{\mathrm{int}}(\Delta t)
	R_z\left(\frac{\pi}{2}\right)
	U_2,
\end{equation}
where $U_2$ denotes the balanced beam-splitter operation generated by
the quarter-revival dynamics and
$U_{\mathrm{int}}(\Delta t)$ describes the nonlinear evolution during
the interaction interval $\Delta t$. The phase imprint
$R_z(\pi/2)$ is applied to one of the clouds at $t=T_r/4$. In the
present protocol, this imprint ensures that the subsequent
intermediate fractional-revival evolution produces two spatially
separated clouds at $t=3T_r/8$, rather than four clouds in its absence. Also, this imprint is responsible for getting $50:50$ split clouds at $t=T_r/2$, given the $50:50$ splitting at $t=T_r/4$.

Including the rotation of the minimum-noise quadrature, phase
encoding, and final readout, the complete protocol can be expressed as
\begin{equation}
	U_{\mathrm{tot}}
	=
	U_2
	R_z(\Phi)
	U_1
	P(\alpha)
	U_1
	U_{\mathrm{SS}}
\end{equation}
The combination $	U_1P(\alpha)U_1=U_{\mathrm{BS}}(\alpha)$
is the generalized beam-splitter discussed in Eq.~\eqref{GBS} and $U_2=U^2_1$ is the balanced beam splitter. We choose $\alpha=\alpha_{\mathrm{min}}-\frac{\pi}{2}$, such that the generalized beam-splitter rotates the minimum-noise
quadrature by $-\alpha_{\mathrm{min}}$ and aligns it with the phase quadrature $J_y$. The resulting state at $t=T_r/2$ is therefore phase-squeezed, allowing the unknown phase $\Phi$ to be encoded with reduced phase variance. The subsequent quarter revival maps this phase information onto the population imbalance for readout.

\begin{figure}[h!]
	\centering
	\includegraphics[width=15 cm]{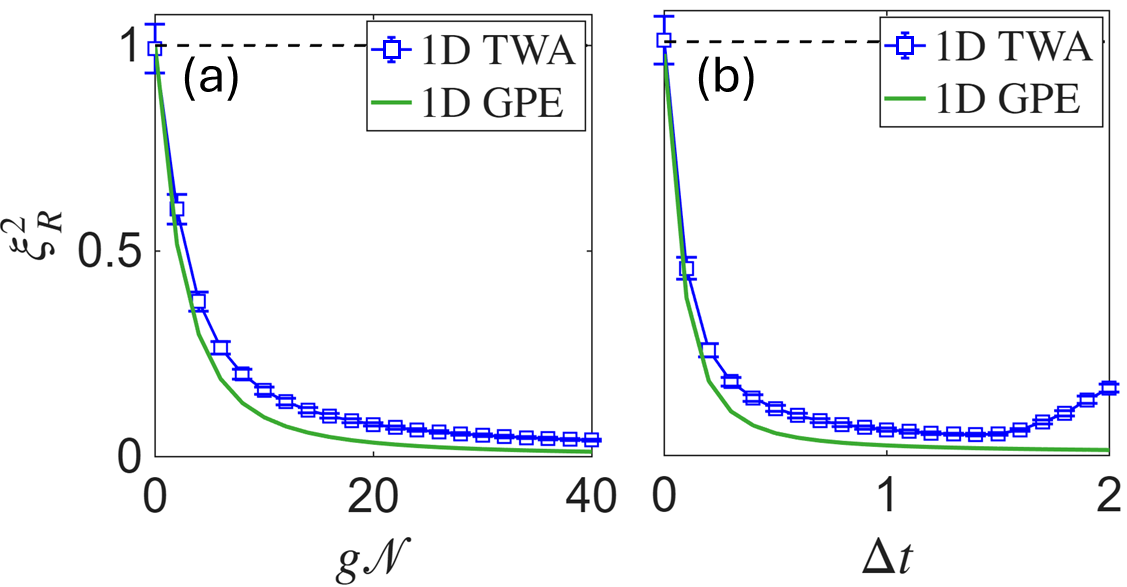}
	\caption{Wineland squeezing parameter $\xi_{R}^2$ with respect to (a) nonlinearity $\mathrm{g}\mathcal{N}$ for fixed quench duration $\Delta t=1$ and (b) quench duration $\Delta t$ for fixed nonlinearity $\mathrm{g}\mathcal{N}=25$. In both cases, the atom number is taken as $\mathcal{N}=10000$. $\xi_{R}^2$ obtained from analytical formula (green solid-line), whose model parameters determined from 1D GPE  is compared with the one obtained through TWA simulation (blue squares). Here, time $t$ is in the units of $\omega_x^{-1}$.} 
	\label{analytical}
\end{figure}
\subsection{Results}
Based on the above protocol, we obtain optimal spin squeezing along an axis oriented at an angle $\alpha_{\mathrm{min}}$ with respect to the $J_y$ axis. To quantify the amount of squeezing along this optimal direction, we calculate the Wineland parameter $\xi_R^2$ as a function of the nonlinearity $\mathrm{g}\mathcal{N}$ for different atom
numbers $\mathcal{N}$. In the linear-squeezing limit, we obtain the Wineland squeezing parameter as Eq.~\eqref{eq:xiR2_linear}:
\begin{equation}
	\xi_{R,\mathrm{lin}}^2=1-\frac{\mathcal{N}\Lambda}{2}
	\left[
	\sqrt{4+\mathcal{N}^2\Lambda^2}-\mathcal{N}\Lambda
	\right].\label{xi_lin}
\end{equation}
where $\Lambda=\int\limits_{0}^{\Delta t}\mathrm{d}t'\chi_{\mathrm{eff}}(t')$ is the squeezing degree. The Wineland parameter is also obtained from the TWA simulation and compared with the $\xi_{R,\mathrm{lin}}^2$ computed from the above analytical model with and determined from 1D GPE simulations, as shown in Fig.~\ref{analytical}. It is evident from Fig.\ref{analytical}(b) that the analytical result is closer to TWA results for small quench times $\Delta t<1.5$, whereas for $\Delta t>1.5$, there is a disagreement between the analytical and TWA results. This is attributed to the fact that the two-model model fails to capture the multimode dynamics as a result of mixing of two modes. 
 
\begin{figure*}[h!]
	\centering
	\includegraphics[width=16 cm]{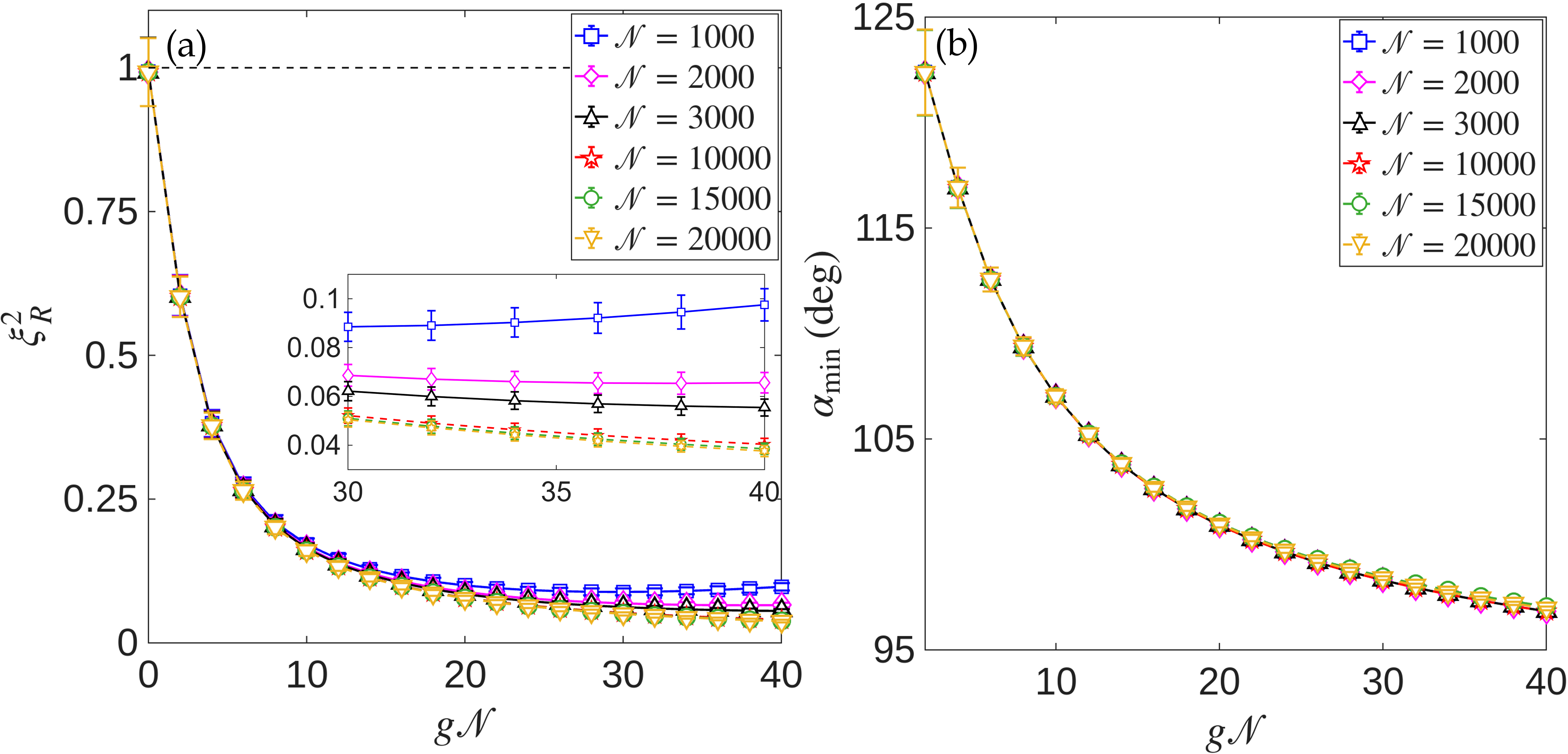}
	\caption{(a) Wineland spin-squeezing parameter $\xi_{R}^2$ as a function of nonlinearity $\mathrm{g}\mathcal{N}$ for different atom numbers $\mathcal{N}$. The inset (b) shows the zoomed in portion near nonlinearity range $\mathrm{g}\mathcal{N}=20-30$ to show the deviations in $\xi_{R}^2$ for small and big $\mathcal{N}$. (c) Corresponding principal squeezing angle $\alpha_{\mathrm{min}}$ in degree with respect to $\mathrm{g}\mathcal{N}$ for different $\mathcal{N}$. Here, the quench duration is taken as $\Delta t=1$ in the units of $\omega_{x}^{-1}$.} 
	\label{OAT_1}
\end{figure*}
The variation of $\xi_R^2$ with $\mathrm{g}\mathcal{N}$ for a quench duration of $\Delta t=1$ is shown in Fig~\ref{OAT_1}(a). We observe a steady decrease in $\xi_R^2$ with increasing $\mathrm{g}\mathcal{N}$, followed by an approximately saturated behavior at the larger nonlinearities considered. The dependence of $\xi_R^2$ on the atom number is almost negligible at low nonlinearities, while a small deviation becomes visible at higher nonlinearities. The inset shows the corresponding deviation in $\xi_R^2$.

\begin{figure*}[ht!]
	\centering
	\includegraphics[width=16 cm]{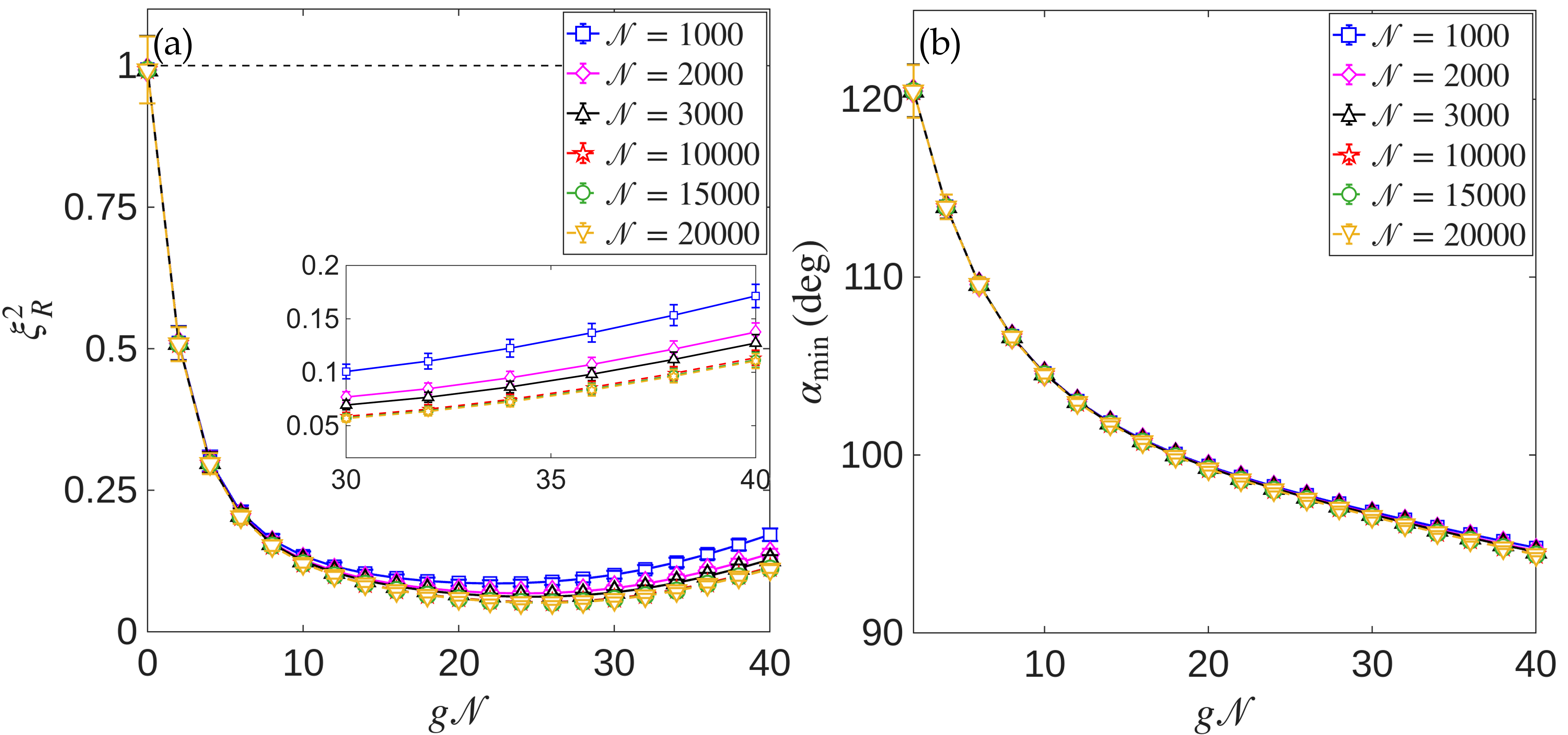}
	\caption{(a) Wineland spin-squeezing parameter $\xi_{R}^2$ as a function of nonlinearity $\mathrm{g}\mathcal{N}$ for different atom numbers $\mathcal{N}$. The inset (b) shows the zoomed in portion near nonlinearity range $\mathrm{g}\mathcal{N}=20-40$ to show the deviations in $\xi_{R}^2$ for small and big $\mathcal{N}$. (c) Corresponding principal squeezing angle $\alpha_{\mathrm{min}}$ in degree with respect to $\mathrm{g}\mathcal{N}$ for different $\mathcal{N}$. Here, the quench duration is taken as $\Delta t=1.5$ in the units of $\omega_{x}^{-1}$.} 
	\label{OAT_2}
\end{figure*}
\begin{figure*}[ht!]
	\centering
	\includegraphics[width=16 cm]{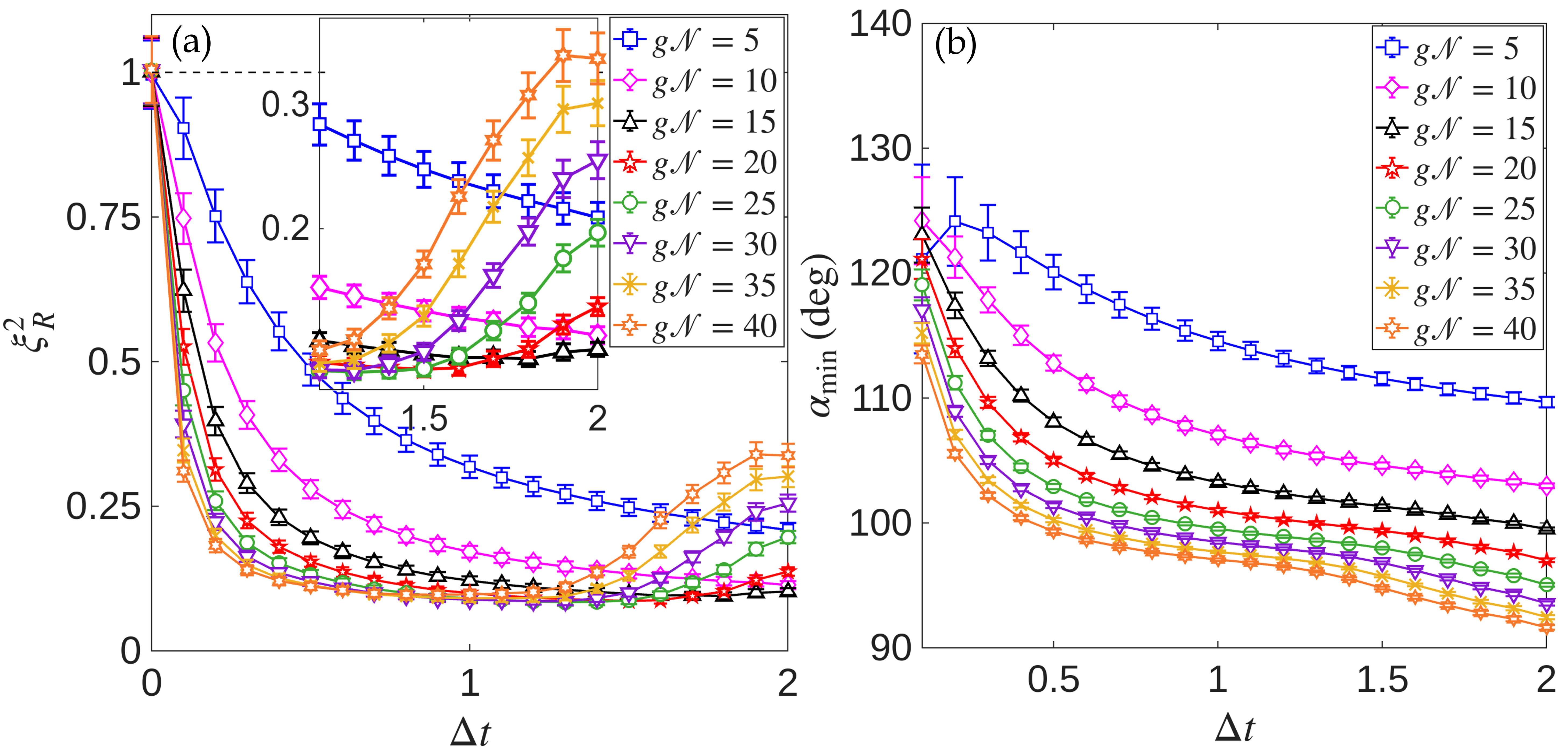}
	\caption{(a) Wineland spin-squeezing parameter $\xi_{R}^2$ as a function of quench duration $\Delta t$ for different nonlinearities $\mathrm{g}\mathcal{N}$. The inset shows the zoomed in portion near duration range $\Delta t=1-2$. (b) Corresponding principal squeezing angle $\alpha_{\mathrm{min}}$ in degree with respect to $\Delta t$ for various $\mathrm{g}\mathcal{N}$. The number of atoms is considered to be $\mathcal{N}=1000$. Here, time is in the units of $\omega_x^{-1}$.} 
	\label{OAT_3}
\end{figure*}
\begin{figure*}[ht!]
	\centering
	\includegraphics[width=16 cm]{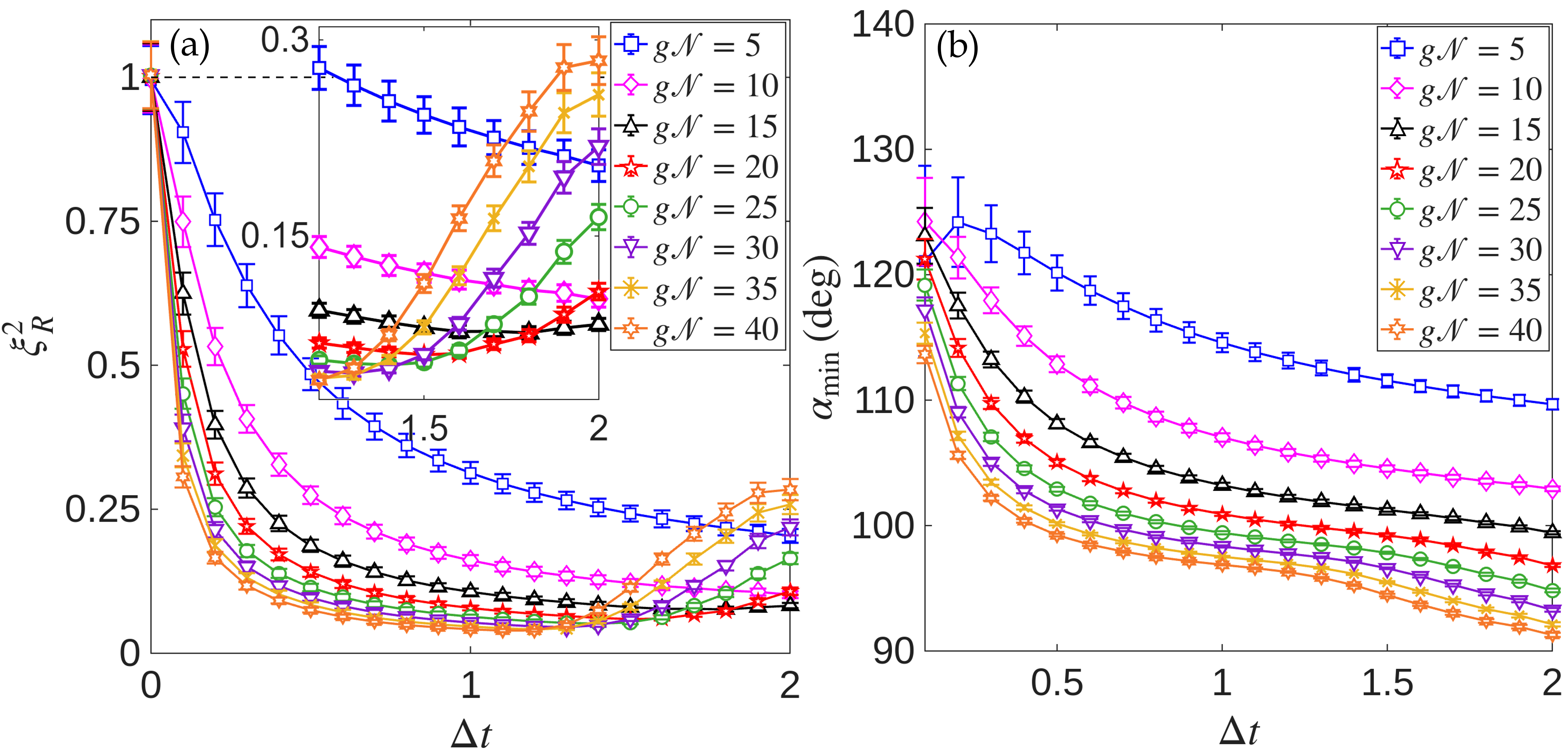}
	\caption{(a) Wineland spin-squeezing parameter $\xi_{R}^2$ as a function of quench duration $\Delta t$ for different nonlinearities $\mathrm{g}\mathcal{N}$. The inset shows the zoomed in portion near duration range $\Delta t=1-2$. (b) Corresponding principal squeezing angle $\alpha_{\mathrm{min}}$ in degree with respect to $\Delta t$ for various $\mathrm{g}\mathcal{N}$.  The number of atoms is considered to be $\mathcal{N}=10000$. Here, time is in the units of $\omega_x^{-1}$.} 
	\label{OAT_4}
\end{figure*}

The principal squeezing angle $\alpha_{\mathrm{min}}$ corresponding to the optimal spin squeezing is shown in Fig.~\ref{OAT_1}(c). With increasing $\mathrm{g}\mathcal{N}$, $\alpha_{\mathrm{min}}$ decreases steadily and approaches $95^\circ$, which is only $5^\circ$ away from the number-squeezed direction at $90^\circ$. The atom number has no significant effect on the principal squeezing angle over the range of nonlinearities considered here.

The choice of quench duration $\Delta t$ also plays an important role in the preparation of the optimally squeezed state at $t=T_r/4+\Delta t$. This is evident from the behavior of $\xi_R^2$ as a function of $\mathrm{g}\mathcal{N}$ for a slightly longer quench duration, $\Delta t=1.5$. The corresponding results are shown in Fig.~\ref{OAT_2}. In this case, $\xi_R^2$ exhibits a gradual increase at higher
nonlinearities, as highlighted in the inset. The increase is more pronounced for smaller atom numbers than for larger atom numbers. On the other hand, $\alpha_{\mathrm{min}}$ approaches much closer to $90^\circ$, corresponding to the number-squeezed direction, compared with the $\Delta t=1$ case.

To further understand the dependence of spin squeezing on the quench duration, we study the Wineland parameter as a function of $\Delta t$ for different nonlinearities $\mathrm{g}\mathcal{N}$. Figure~\ref{OAT_3} summarizes the behavior of $\xi_R^2$ and $\alpha_{\mathrm{min}}$ as a function of $\Delta t$ for different values of $\mathrm{g}\mathcal{N}$. Here,
the atom number is fixed at $\mathcal{N}=1000$. As expected from the OAT picture, the interaction strength determines the amount of spin squeezing, with lower nonlinearities producing weaker squeezing than higher nonlinearities. More importantly, for higher nonlinearities, $\xi_R^2$ initially decreases with increasing $\Delta t$, reaches a minimum, and subsequently increases as the quench duration is increased further. The onset of this increase occurs at smaller $\Delta t$ for larger nonlinearities, as shown in the inset of Fig.~\ref{OAT_3}(a).

The corresponding behavior of $\alpha_{\mathrm{min}}$ shows that the increase in $\xi_R^2$ is accompanied by a shift of the principal squeezing axis toward $90^\circ$, with this effect becoming more pronounced at higher nonlinearities. A similar dependence of $\xi_R^2$ and $\alpha_{\mathrm{min}}$ on $\Delta t$ is observed for the larger atom number $\mathcal{N}=10000$ as shown in Fig.~\ref{OAT_4}. However, the larger atom
numbers yield smaller values of $\xi_R^2$ compared with the smaller atom numbers. For example, at $\mathrm{g}\mathcal{N}=40$, the Wineland parameter reaches $\xi_R^2=0.038\pm0.003$ for
$\mathcal{N}=20000$, compared with $\xi_R^2=0.065\pm0.004$ for
$\mathcal{N}=2000$, with $\Delta t=1$ in both cases.

The increase in the Wineland parameter beyond an optimal quench duration is particularly interesting and is associated with the onset of multimode dynamics. We observe that the increase in $\xi_R^2$ occurs when the two BEC clouds begin to overlap during the nonlinear evolution. This overlap occurs at shorter times for larger nonlinearities, consistent with the earlier onset of the increase in $\xi_R^2$ at larger $\mathrm{g}\mathcal{N}$. Once the two clouds begin to overlap, the system begins to deviate from the ideal two-mode picture, with additional spatial-mode contributions becoming increasingly important. The resulting multimode dynamics modify the spin distribution and cause the initially generated squeezing to decrease. So, this gives us the optimal range of $\Delta t$ within which we have maximum squeezing.

\begin{figure*}[ht!]
	\centering
	\includegraphics[width=15 cm]{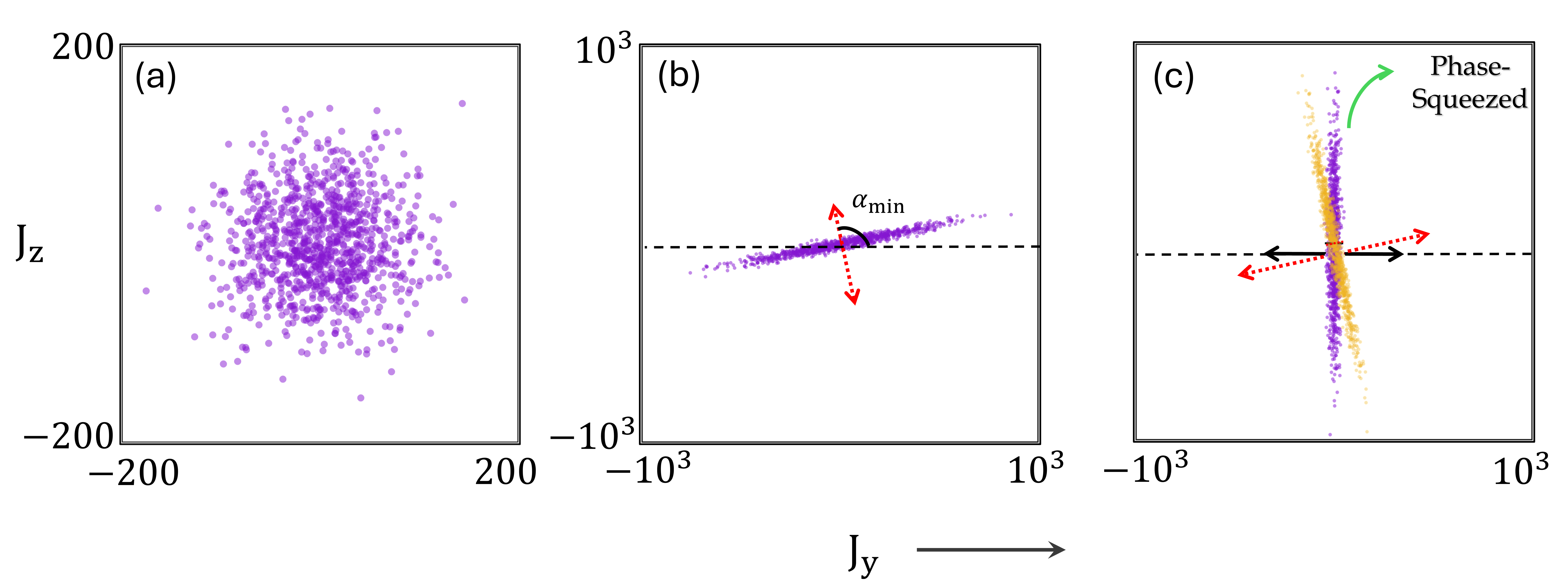}
	\caption{Collective spin distribution in the $J_y-J_z$ transverse plane and the rotation of the minimum noise axis towards the $J_{y}$-axis. (a) The coherent spin state at $t=T_r/4$ followed by (b) the spin-squeezed state (SSS) along the optimal axis at an angle $\alpha_{\mathrm{min}}$ with respect to the $J_y$-axis at $t=T_r/4+\Delta t$. (c) A SSS obtained at $t=T_r/2$ with its minimum noise axis directed at an angle $\alpha_{\mathrm{min}}-\pi/2$ with the $J_y$-axis. A phase-squeezed state is also shown in the same figure (c) which is obtained as a result of modified quarter-revival evolution to $t=T_r/2$. The red dotted-line indicates an axis passing through the minimum noise direction of the SSS, whereas the black solid line denotes the axis passing through the minimum noise axis of the phase-squeezed state.} 
	\label{SS_rot}
\end{figure*}

\subsubsection{Noise axis rotation via Generalized Beam-Splitter} 
The spin-squeezed state obtained by the interaction-quench FR protocol generates a much stronger squeezing compared to that obtained through nonlinear quarter revival evolution discussed in the previous section. Wineland parameter as small as $\xi_{R}^2=0.0349\pm 0.0021$ is obtained for $\mathrm{g}\mathcal{N}=40$ with $\mathcal{N}=20000$, with a quench duration of $\Delta t=1.2$. However, the next step is to rotate this optimal squeezed quadrature into the phase-sensitive direction along $J_y$. This phase-squeezed state can be used to encode the interferometric phase, which is converted into a number-squeezed state from which population difference gives the phase information with reduced fluctuations. 

The state at $t=T_r/4$, after a phase imprint of $\pi/2$, is represented by a circular distribution in the $J_y-J_z$ plane as shown in Fig.~\ref{SS_rot}(a). When such a state undergoes interaction quench of strength $g$, for a duration $\Delta t$, results in a spin-squeezed state (Fig.~\ref{SS_rot}(b)) with the minimum noise direction oriented at an angle $\alpha_{\mathrm{min}}$ with the $J_y$ axis. Under a conventional quarter-revival evolution $U_2$, which corresponds to a balanced beam splitter, one would get a spin-squeezed state further rotated by a $90^\circ$ at $t=T_r/2$. However, under the modified quarter-revival $U_1P(\alpha_{\mathrm{min}}-\pi/2)U_1$, which corresponds to generalized beam splitter, we get a phase-squeezed state at the end of $t=T_r/2$. Both these spin-squeezed state at an angle $\alpha_{\mathrm{min}}-\pi/2$ with $J_y$-axis and the phase-squeezed state are shown in Fig.~\ref{SS_rot}(c). It is worth mentioning that the modified quarter-revival $U_1P(\alpha_{\mathrm{min}}-\pi/2)U_1$ operation does not result in any population imbalance as the $\pi/2$ phase imprint at $T_r/4$ changes the relative phase between the $50:50$ split BEC from $\pi/2$ to $0$$\pmod{\pi}$. However, it being an $X-$rotation, rotates the spin ellipse while keeping the mean spin vector unaffected. 

\subsubsection{Below-SQL Phase sensitivity}
With the phase-squeezed state at $t=T_r/2$, an unknown interferometric phase $\Phi$ is imprinted to one of the clouds. We numerically imprint such a phase using the operation $\exp(i\Phi M(x))$ on the wavefunction, where $M(x)$ is a mask function given by:
\begin{eqnarray}
	M(x)=\frac{1}{2}\left(1-\tanh\left(\frac{|y|-2\pi}{\sigma}\right)\right) 
\end{eqnarray}
where $\sigma$ is the mask width and $|y|=\bmod(x+C/2,C)-C/2$ with $C=2\pi R$ being the circumference of the ring. Similar technique has been employed for imprinting phases at $t=T_r/4$ and $3T_r/8$. The interferometric phase $\Phi$ corresponds to an unknown phase, which is the relative phase accumulated between the two clouds as a result of an external perturbation or gauge fields. For instance, one could think of gravimetry using the FR interferometer, where the ring is briefly tilted at $T_r/2$ such that a phase of $\Phi\approx 2\bar{g}R\tau\sin\Omega$ is accumulated between the two clouds, where $\bar{g}=g/{a_x\omega^2_{x}}$and $g$ is the acceleration due to gravity~\cite{kialka2020orbital}. The amount of tilt is given by $\Omega$ and $\tau$ denotes the time duration of tilt. 
\begin{table}[htbp]
	\centering
	\caption{Comparison of the phase sensitivity $\Delta\Phi$ obtained
		using the linear-fit and central-difference (CD) estimates of
		$\partial\langle J_z\rangle/\partial\Phi$, together with the Wineland
		prediction $\Delta\Phi_{\mathrm{W}}=\xi_R/\sqrt{\mathcal{N}}$, for
		different atom numbers $\mathcal{N}$. Here, the nonlinearity is taken as $\mathrm{g}\mathcal{N}=40$ and the quench duration as $\Delta t=1$ in the units of $\omega_{x}^{-1}$.}
	\label{tab1}
	\begin{tabular}{c c c c}
		\toprule
		$\mathcal{N}$
		& $\Delta\Phi_{\mathrm{fit}}\times10^{3}$
		& $\Delta\Phi_{\mathrm{CD}}\times10^{3}$
		& $\Delta\Phi_{\mathrm{W}}\times10^{3}$ \\
		\midrule
		$2000$  & $5.706$ & $5.706$ & $5.686$ \\
		$4000$  & $4.265$ & $4.265$ & $4.255$ \\
		$6000$  & $2.722$ & $2.722$ & $2.718$ \\
		$8000$  & $2.292$ & $2.292$ & $2.289$ \\
		$10000$ & $2.015$ & $2.015$ & $2.012$ \\
		\bottomrule
	\end{tabular}
\end{table}

Given the interferometric phase encoded in the phase-squeezed state, the following quarter-revival evolution rotates this state into a number-squeezed state at $t=3T_r/4$. A measurement of population imbalance between the two arms $a$ and $b$ of the interferometer gives the phase information. We numerically execute the entire protocol discussed in Fig.~(\ref{sch_SS}) by scanning the interferometric phase around its operational point $\Phi=0$, at $t=T_r/2$. Following this, we use the error propagation formula to estimate the phase sensitivity, which is given by~\cite{martin2012quantum,haine2018quantum,szigeti2020high}: 
\begin{equation}
	\Delta\Phi=\sqrt{\text{Var}(J_z)}\left|\frac{\partial\langle J_z\rangle}{\partial\Phi}\right|^{-1}
\end{equation}
Here, the slope $\frac{\partial\langle J_z\rangle}{\partial\Phi}$ is obtained by two methods: (1) from a linear fit in the $\langle J_z\rangle$ vs $\Phi$ plot and (2) from the central-difference method. We used two methods to verify if the slope obtained is internally consistent. 

For a given nonlinearity $\mathrm{g}\mathcal{N}=40$ quenched for a duration of $\Delta t=1$, the table~\ref{tab1} compares the phase sensitivities
obtained using the two independent estimates of the phase response,
namely, the linear-fit and central-difference estimates of
$\partial\langle J_z\rangle/\partial\Phi$. The two approaches yield identical values to the precision shown, demonstrating that the estimated phase response is insensitive to the choice of slope
estimation method. Furthermore, the directly evaluated phase sensitivities are in excellent agreement with the Wineland prediction
$\Delta\Phi_{\mathrm{W}}=\xi_R/\sqrt{\mathcal{N}}$. This agreement confirms that the spin squeezing generated during the interaction quenched stage translates into the expected sub-shot-noise phase
sensitivity in the complete interferometric sequence.
\begin{figure}[ht!]
	\centering
	\includegraphics[width=10 cm]{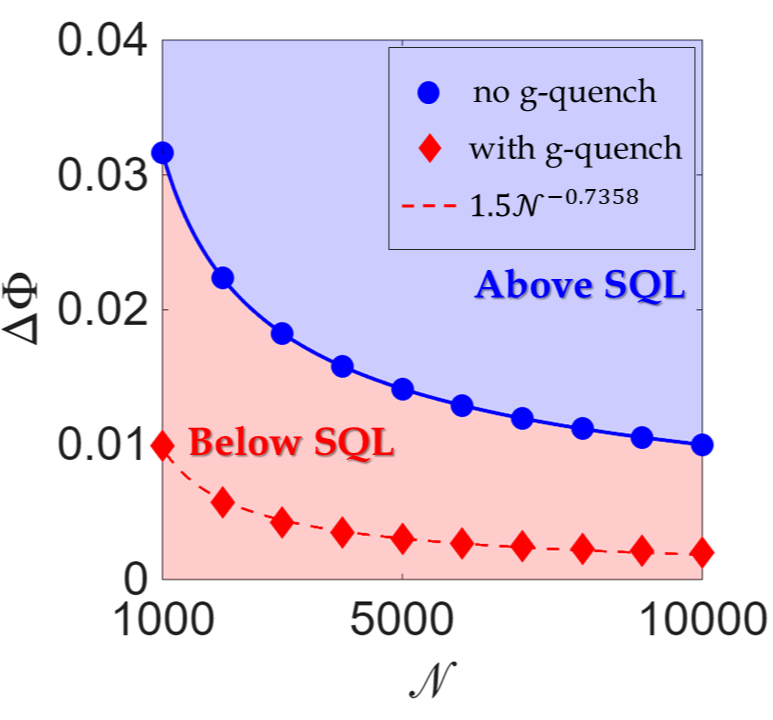}
	\caption{Variation of phase sensitivity $\Delta\Phi$ with respect to the atom number $\mathcal{N}$. The no interaction quench corresponding to SQL is denoted by circles  and the solid line denotes the SQL boundary given by ${\mathcal{N}}^{-0.5}$. On the other hand, the sub-SQL sensitivity due to the interaction quench is depicted by diamonds and the corresponding dashed line denotes the $1.5\mathcal{N}^{-0.7358}$ fit.} 
	\label{SSQL}
\end{figure}

For the same nonlinearity as above $\mathrm{g}\mathcal{N}=40$ quenched for a duration of $\Delta t=1$, the phase sensitivity is plotted as a function of the atom number $\mathcal{N}$ in Fig.~\ref{SSQL}. In the absence of any interaction quench, we obtain the phase sensitivity scaling as $1/\sqrt{\mathcal{N}}$, illustrated by the circles in the figure. The blue-shaded region highlights the above SQL regime whereas the red region highlights the below SQL regime. In the presence of interaction quench at $t=T_r/4$, we obtain the sub-SQL phase sensitivity indicated by the diamond markers. We supposedly fit our below SQL data with nonlinear function $\Delta\Phi=1.5\mathcal{N}^{-0.7358}$. The fit indicates that the phase sensitivity lies at almost halfway between SQL and the Heisenberg limit which scales as $1/\mathcal{N}$.

\begin{figure}[ht!]
	\centering
	\includegraphics[width=12 cm]{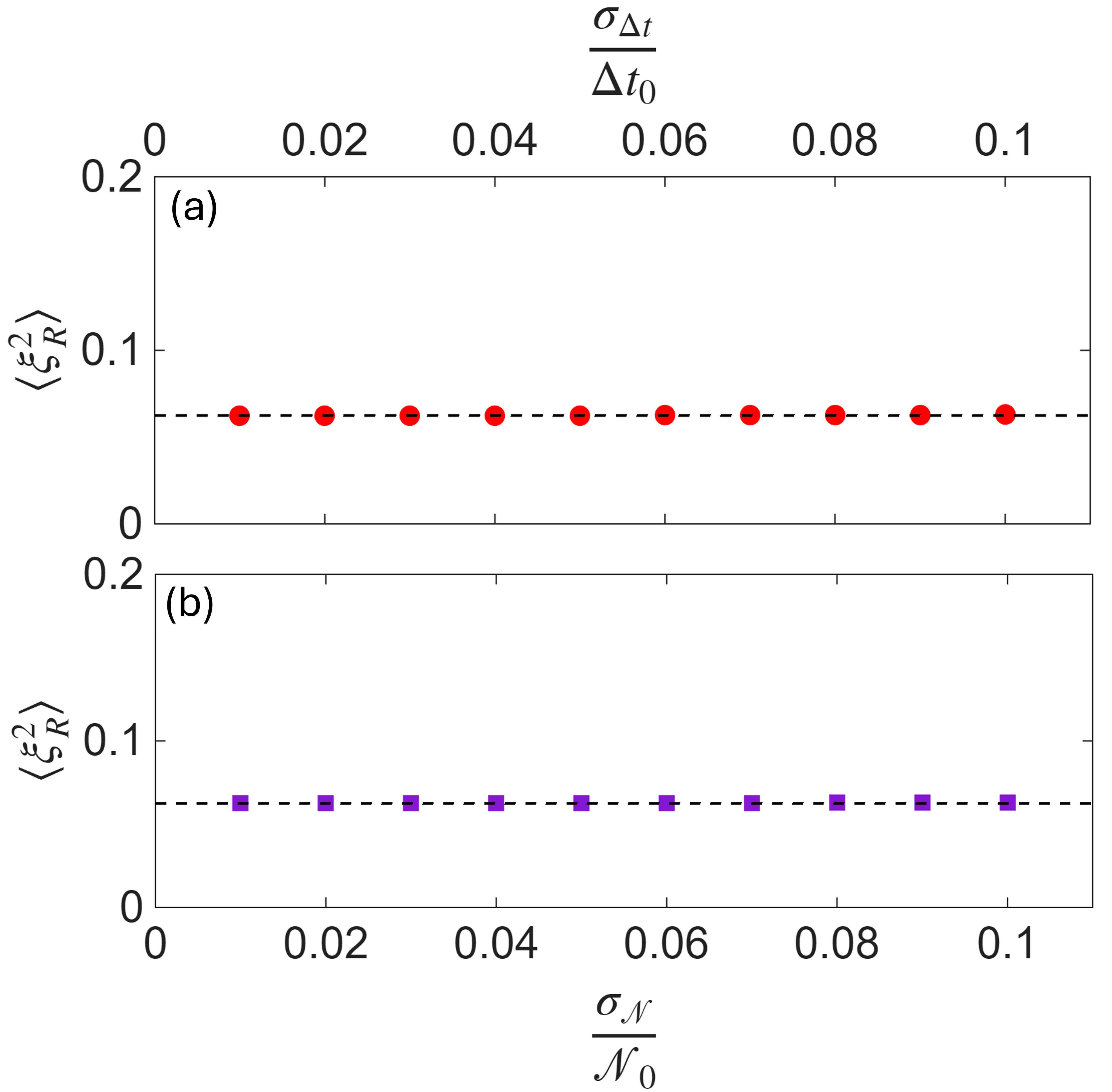}
	\caption{The effect of (a) Gaussian shot-to-shot  pulse timing fluctuations of variance $\sigma_{\Delta t}$, corresponding to the interaction quench $\Delta t$ and (b) Gaussian shot-to-shot atom number fluctuations of variance $\sigma_{\mathcal{N}}$. Here, we consider  $\mathcal{N}=10000$, $g\mathcal{N}=25$, $\Delta t=1$ in the units of $\omega_{x}^{-1}$.} 
	\label{Robustness}
\end{figure}
\subsection{Robustness of Squeezing}
The generation of metrologically useful squeezing requires not only a favorable nominal operating point but also sufficient robustness against experimental fluctuations~\cite{szigeti2020high}. In order to determine the robustness of our protocol, we discuss two experimental imperfections: \textit{i}) Pulse timing imperfections, and \textit{ii}) Atom number fluctuations.
\subsubsection{Pulse timing imperfection}
In an experiment, the duration of nonlinear interaction pulse $\Delta t$ may vary from shot-to-shot because of finite timing resolution in the switching electronics and other imperfections in
the pulse sequence~\cite{cheinet2008measurement}. We model this by allowing the pulse duration to
fluctuate around its nominal value, $\Delta t = \Delta t_0 + \eta$, where $\eta$ is a zero-mean Gaussian random variable, while holding the nonlinearity $g\mathcal{N}$ and the atom number $\mathcal{N}$ fixed. For each value of the timing jitter, we compute the Gaussian-weighted average of the per-realization Wineland squeezing parameter over the resulting distribution of pulse durations. The resulting mean Wineland parameter $\langle \xi_{R}^2\rangle$ as a function of timing uncertainty $\sigma_{\Delta t}/\Delta t_0$ is shown in Fig.~\ref{Robustness}(a). The average squeezing remains nearly constant over the considered range, with only a slight increase as the timing uncertainity is increases.  
\subsubsection{Shot-to-shot atom-number fluctuations}
We have shown above that the degree of metrological squeezing depends on the nonlinearity parameter $g\mathcal{N}$ at fixed atom number $\mathcal{N}$. In practice, however, $\mathcal{N}$ fluctuates from shot to shot, and such fluctuations modify the effective nonlinearity, thereby affecting both the degree of phase-space shearing and the orientation of the squeezed quadrature~\cite{szigeti2020high}. So, we model the atom number as fluctuating around its nominal value $\mathcal{N}_0$ according to a Gaussian distribution. Here also we calculate $\langle\xi_R^2\rangle$ as a function of uncertainty fraction $\sigma_{\mathcal{N}}/\mathcal{N}_0$, shown in Fig.~\ref{Robustness}(b). The averaged squeezing changes only weakly over the considered range of atom-number fluctuations. In particular, even for relative fluctuations as large as $10\%$, the Wineland parameter remains close to its nominal value. This indicates that the squeezing generated by our protocol is insensitive to moderate shot-to-shot variations in the atom number.

\section{Conclusion:}\label{s5_c7}

We have investigated the fractional-revival (FR) dynamics of a matter-wave packet in a 1D ring and demonstrated its application to quantum-enhanced matter-wave interferometry. The central result of this work is that FR dynamics can provide both the beam-splitter and recombination operations, while controlled nonlinear evolution between these operations can be used to generate spin squeezing and enhance interferometric sensitivity. In particular, the quarter revival provides a coherent $50{:}50$ splitting of the matter wave, acting as a generalized beam splitter without the need for an optical lattice or gaussian barrier as used in soliton and droplet interferometers\cite{martin2012quantum,raghav2025quantum}.

Within the mean-field description, we showed that interactions modify the revival dynamics, with increasing nonlinearity reducing the revival time. We then incorporated quantum fluctuations using the Truncated-Wigner Approximation (TWA). The condensate fraction exhibits decay accompanied by partial revivals at FR instances, whereas the phase coherence undergoes a predominantly steady decay with the increasing nonlinearity. The loss of phase coherence becomes more pronounced for smaller atom numbers and stronger nonlinearities, thereby identifying the regime in which the FR beam splitter remains suitable for interferometry. Moreover, for atom numbers as high as $\mathcal{N}=10000$, $15000$, and $20000$, the first-order inter-cloud correlation $c_{12}$ and relative phase coherence $|\langle e^{i\phi_{12}}\rangle|$ remain effectively unity over the range of nonlinearities considered.

The nonlinear evolution up to the quarter revival generates moderate spin squeezing. Although the number-squeezing parameter $\xi_{\mathcal{N}}^2$ and the spin-squeezing parameter $\xi_{\mathcal{S}}^2$ increase from their SQL values when evaluated along the laboratory $J_z$ axis, the covariance matrix reveals squeezing along an optimally rotated direction. The corresponding Wineland parameter $\xi_R^2$ reaches approximately $-2$ dB of spin squeezing for $\mathrm{g}\mathcal{N}=3$. Further nonlinear evolution eventually degrades this squeezing, as evidenced by the increase of $\xi_R^2$ at $t=T_r/2$ for the same nonlinearities.

The main result of this work is a quantum-enhanced FR interferometry protocol in which the interaction is introduced as a controlled quench after the initial beam-splitter operation. In this protocol, a noninteracting localized condensate is first prepared and evolved to the quarter revival, where the FR dynamics produces the initial $50{:}50$ beam splitter. The interaction is then switched on for a controlled duration $\Delta t$, generating spin squeezing through nonlinear evolution, after which the subsequent FR dynamics provides the recombination. By tuning the interaction strength $g$ and the quench duration $\Delta t$, we obtain a minimum Wineland parameter corresponding to spin squeezing of $-14.57\pm0.27$ dB. For a given nonlinearity, $\xi_R^2$ begins to increase with increasing quench duration $\Delta t$, which is associated with the increasing involvement of multimode dynamics. Nevertheless, by tuning $g$ and $\Delta t$, an optimal reduction of the noise variance below the SQL can be achieved.

The proposed protocol is compatible with experimentally accessible
parameters. We consider a quasi-1D geometry with
$\omega_x=2\pi\times50$ Hz and $\omega_r=2\pi\times500$ Hz,
giving $\lambda=0.1$, and use ${}^{39}\mathrm{K}$ atoms, which
exhibit broad Feshbach resonances~\cite{herbst2022rapid,chin2010feshbach}
that provide control over the interaction strength. As demonstrated
by Petrucciani et al.~\cite{petrucciani2026mach}, the attractive
magnetic dipolar contribution can be compensated by a small,
oppositely signed $s$-wave elastic contribution, allowing an
effectively noninteracting condensate to be realized. This
interaction control is particularly important for our second
protocol, where a noninteracting condensate is required for the
initial fractional-revival beam-splitter operation, after which the
interaction is introduced as a controlled quench to generate
squeezing.

For the TWA simulations, we use $M=144$ BdG modes to construct the initial TWA ensemble and $N_{\mathrm{traj}}=1000$ stochastic trajectories. The use of a finite BdG-mode basis and the associated convergence requirements are consistent with the established validity considerations of the truncated-Wigner method for Bose-condensed gases!\cite{sinatra2002truncated,blakie2008dynamics}. Both parameters are chosen based on systematic convergence tests to ensure that the reported results are numerically converged. The $x$-coordinate is discretized into 1024 grid points with a step size of $\Delta x=0.03$, while the time step is $\Delta t=0.003$. The spatial and temporal discretization follows standard numerical treatments of the Gross–Pitaevskii equation, including spectral/time-splitting approaches~\cite{bao2003numerical}. In imaginary time propagation, the relative change in the energy functional and chemical potential are monitored for convergence~\cite{muruganandam2009fortran,bao2004computing}, and the ground state is accepted once an absolute tolerance of $10^{-10}$ is achieved.

Overall, our results establish FR dynamics as a framework for quantum-enhanced matter-wave interferometry. The FR dynamics provides both the beam-splitter and recombination operations, while the controlled interaction quench generates spin squeezing between these operations within a single-component condensate.The demonstrated $-14.57\pm0.27$ dB spin squeezing, together with experimentally accessible interaction control and converged multimode TWA simulations, establishes a feasible route to realizing a strongly squeezed matter-wave interferometer using the intrinsic dynamics of matter waves in a ring waveguide.

\appendix
\section{Analytic Wineland parameter in the effective two-mode one-axis-twisting model}
\label{app:analytic_xiR2}

We estimate the Wineland parameter by mapping the nonlinear dynamics of the two spatial wave-packet modes onto an effective one-axis-twisting (OAT) Hamiltonian~\cite{kitagawa1993squeezed}, assuming negligible spatial overlap between the modes during the nonlinear evolution. Following the approach of Ref.~\cite{szigeti2020high}, this provides an analytic reference for the truncated-Wigner results of Sec.~\ref{s4_c7}.

\subsection{Effective two-mode Hamiltonian}
\label{app:two_mode_hamiltonian}

Within a single-mode-per-window approximation, the field operator is $\hat\psi(x,t)=\psi_a(x,t)\hat a+\psi_b(x,t)\hat b$, where $\hat a$ and $\hat b$ are bosonic operators associated with the disjoint windows $\Xi_a$ and $\Xi_b$ (Eq.~\eqref{window}). The modes satisfy $\int\mathrm{d}x\,|\psi_m|^2=1$ ($m=a,b$) and $\int\mathrm{d}x\,\psi_a^*\psi_b\simeq0$. Inserting this into the contact interaction $\hat H_{\mathrm{int}}=\frac{g(t)}{2}\int\mathrm{d}x\,\hat\psi^{\dagger}\hat\psi^{\dagger}\hat\psi\hat\psi$, the intermode terms vanish by the negligible overlap (i.e.\ $\chi_{ab}\to0$), leaving
\begin{equation}
	\hat H_{\mathrm{int}}(t)=\sum_{m=a,b}\chi_m(t)\,\hat{\mathcal N}_m(\hat{\mathcal N}_m-1),
	\qquad
	\chi_m(t)=\frac{g(t)}{2}\int\mathrm{d}x\,|\psi_m(x,t)|^4,
	\label{eq:two_mode_interaction}
\end{equation}
with $\hat{\mathcal N}_m$ the number operator of mode $m$ (energy in the units of $\hbar\omega_x$). Defining $\hat J_z=(\hat{\mathcal N}_a-\hat{\mathcal N}_b)/2$ and fixing the total atom number $\hat{\mathcal N}=\mathcal N$, we have $\hat{\mathcal N}_{a,b}=\mathcal N/2\pm\hat J_z$, so that
\begin{equation}
	\hat H_{\mathrm{int}}(t)=\bigl[\chi_a+\chi_b\bigr]\hat J_z^2+(\mathcal N-1)\bigl[\chi_a-\chi_b\bigr]\hat J_z+\mathrm{const.}
	\label{eq:H_spin_full}
\end{equation}
The term linear in $\hat J_z$ vanishes for $\chi_a=\chi_b$. More generally it commutes with the quadratic term and only rotates the spin about $z$, which can be absorbed into the phase of the subsequent beam splitter~\cite{szigeti2020high}. We therefore retain
\begin{equation}
	\hat H_{\mathrm{OAT}}(t)=\chi_{\mathrm{eff}}(t)\,\hat J_z^2,
	\qquad
	\chi_{\mathrm{eff}}(t)=\chi_a(t)+\chi_b(t).
	\label{eq:H_OAT}
\end{equation}
Since $\hat H_{\mathrm{OAT}}(t)$ at different times is proportional to the same operator, the evolution is $\hat U=\exp[-i\Lambda\hat J_z^2]$ with accumulated nonlinear phase
\begin{equation}
	\Lambda(t)=\int_0^t\mathrm{d}t'\,\bigl[\chi_a(t')+\chi_b(t')\bigr].
	\label{eq:lambda_definition}
\end{equation}

\subsection{Wineland parameter}
\label{app:wineland}

We take the initial state to be a coherent spin state (CSS) polarized along
$+x$, $\ket{j=\mathcal N/2,\,J_x=j}$. Under $\hat U=\exp[-i\Lambda\hat J_z^2]$,
the mean spin remains along $x$, with $\langle\hat J_y\rangle=\langle\hat J_z\rangle=0$,
and the remaining moments are~\cite{kitagawa1993squeezed}
\begin{equation}
	\begin{aligned}
		\langle\hat J_x\rangle
		&=\frac{\mathcal N}{2}\cos^{\mathcal N-1}\Lambda,\\[2pt]
		\operatorname{Var}(\hat J_y)
		&=\frac{\mathcal N}{4}\left[1+\frac{\mathcal N-1}{2}
		\bigl(1-\cos^{\mathcal N-2}2\Lambda\bigr)\right],\\[2pt]
		\operatorname{Var}(\hat J_z)
		&=\frac{\mathcal N}{4},\\[2pt]
		\operatorname{Cov}(\hat J_y,\hat J_z)
		&=\frac{\mathcal N(\mathcal N-1)}{4}\sin\Lambda\,\cos^{\mathcal N-2}\Lambda,
	\end{aligned}
	\label{eq:OAT_moments}
\end{equation}
where $\operatorname{Cov}(\hat J_y,\hat J_z)=\tfrac12\langle\{\Delta\hat J_y,\Delta\hat J_z\}\rangle$
is the symmetrized covariance. The variance of $\hat J_z$ is unchanged
because $[\hat J_z,\hat H_{\mathrm{OAT}}]=0$.

The minimum transverse variance is the smaller eigenvalue of the
$(J_y,J_z)$ covariance matrix,
\begin{equation}
	\Delta J_{\min}^2=\frac12\Bigl[\operatorname{Var}(\hat J_y)+\operatorname{Var}(\hat J_z)
	-\sqrt{\bigl[\operatorname{Var}(\hat J_y)-\operatorname{Var}(\hat J_z)\bigr]^2
		+4\operatorname{Cov}(\hat J_y,\hat J_z)^2}\,\Bigr],
	\label{eq:minimum_variance}
\end{equation}
and the Wineland parameter is
\begin{equation}
	\xi_R^2=\frac{\mathcal N\,\Delta J_{\min}^2}{|\langle\hat{\boldsymbol J}\rangle|^2}
	=\frac{4\,\Delta J_{\min}^2}{\mathcal N\cos^{2(\mathcal N-1)}\Lambda},
	\label{eq:xiR2_exact}
\end{equation}
which equals unity at $\Lambda=0$.

\subsection{Linear-squeezing limit}
\label{app:linear_limit}

For $\mathcal N\to\infty$ and $\Lambda\to0$ at fixed $u=\mathcal N\Lambda$, the moments reduce to $\langle\hat J_x\rangle\to\mathcal N/2$, $\operatorname{Var}(\hat J_y)\to\frac{\mathcal N}{4}(1+u^2)$, $\operatorname{Var}(\hat J_z)\to\frac{\mathcal N}{4}$, and $\operatorname{Cov}(\hat J_y,\hat J_z)\to\frac{\mathcal N u}{4}$. The covariance-matrix eigenvalues are then $\frac{\mathcal N}{8}\bigl[2+u^2\pm u\sqrt{u^2+4}\bigr]$, giving
\begin{equation}
	\xi_{R,\mathrm{linear}}^2=1+\frac{u^2}{2}-\frac{u}{2}\sqrt{u^2+4},
	\qquad u=\mathcal N\Lambda,
	\label{eq:xiR2_linear}
\end{equation}
which reproduces Eq.~\eqref{xi_lin} of the main text. This expression is the linear-squeezing-regime approximation to the finite-$\mathcal{N}$ OAT result. It approaches unity at $\mathcal{N}\Lambda=0$ and agrees with Eq.~\eqref{eq:xiR2_exact} when $\mathcal{N}\Lambda\ll1$.
\bibliographystyle{unsrt}
\bibliography{references}	
\end{document}